\documentclass[journal]{IEEEtran}
\usepackage{amsmath,amsfonts}
\usepackage{algorithmic}
\usepackage{algorithm}
\usepackage{array}
\usepackage[caption=false,font=normalsize,labelfont=sf,textfont=sf]{subfig}
\usepackage{textcomp}
\usepackage{stfloats}
\usepackage{url}
\usepackage{verbatim}
\usepackage{graphicx}
\usepackage{cite}
\usepackage{epstopdf}
\usepackage{diagbox}
\usepackage{booktabs}
\begin{document}

\title{A Secure Splitting and Acceleration Strategy for TCP/QUIC in Interplanetary Networks}

\author{Jianhao~Yu, Ye~Li,~\IEEEmembership{Member,~IEEE}, Qingfang Jiang, Shuai Liu,~\IEEEmembership{Graduate Student Member,~IEEE}, Wenfeng~Li,~\IEEEmembership{Member,~IEEE}, and Kanglian~Zhao,~\IEEEmembership{Member,~IEEE}
\thanks{ \textit{Corresponding author: Kanglian Zhao.}}
\thanks{Jianhao~Yu, Qingfang Jiang, Shuai Liu, Wenfeng~Li, and Kanglian~Zhao are with the School of Electronic Science and Engineering, Nanjing University, Nanjing 210093, China (e-mail: jianhaoyu@smail.nju.edu.cn; jiangqf0406@163.com; liushuai\_nju@foxmail.com; leewf\_cn@hotmail.com; zhaokanglian@nju.edu.cn).}
\thanks{ Ye~Li is with the School of Information Science and Technology, Nantong University, Nantong 226019, China, and also with the Nantong Research Institute for Advanced Communication Technologies (NRIACT), Nantong 226019, China (e-mail: yeli@ntu.edu.cn).}
}


\maketitle

\begin{abstract}
Interplanetary networks (IPNs) present unique challenges such as extreme delay, high loss, and frequent disruptions that severely degrade the performance of conventional transport protocols like Transmission Control Protocol (TCP) and Quick UDP Internet Connection (QUIC). To address these issues, we propose a secure transport-acceleration strategy tailored for IPNs. This strategy is founded on our Non-Transparent Secure Proxy (NTSP) architecture, which enables connection splitting for end-to-end encrypted flows while preserving application-layer security. Based on the NTSP, we design an IPN-aware transport policy that combines (i) a rate-based congestion control algorithm exploiting the pre-scheduled nature of deep-space links to achieve stable, efficient bandwidth utilization, and (ii) an adaptive packet-level forward error correction scheme to provide low-latency loss recovery without retransmissions. Furthermore, we introduce a theoretically-grounded backpressure flow control mechanism, deriving an analytical model for optimal buffer sizing to mitigate rate mismatch and prevent bufferbloat. The strategy is implemented in a prototype system, PEPspace, and evaluated in representative Earth–Moon scenarios. Results show near-capacity, stable goodput and substantially improved delivery performance compared with TCP/QUIC variants and existing Performance Enhancing Proxies, while maintaining low latency and robust data delivery across intermittent links. The NTSP architecture is further discussed as a foundational framework for future unified IP/DTN architectures, bridging a key architectural gap in heterogeneous space networks.
\end{abstract}


\section{Introduction}
\begin{figure}[t]
    \centering
    \includegraphics[width=0.485\textwidth]{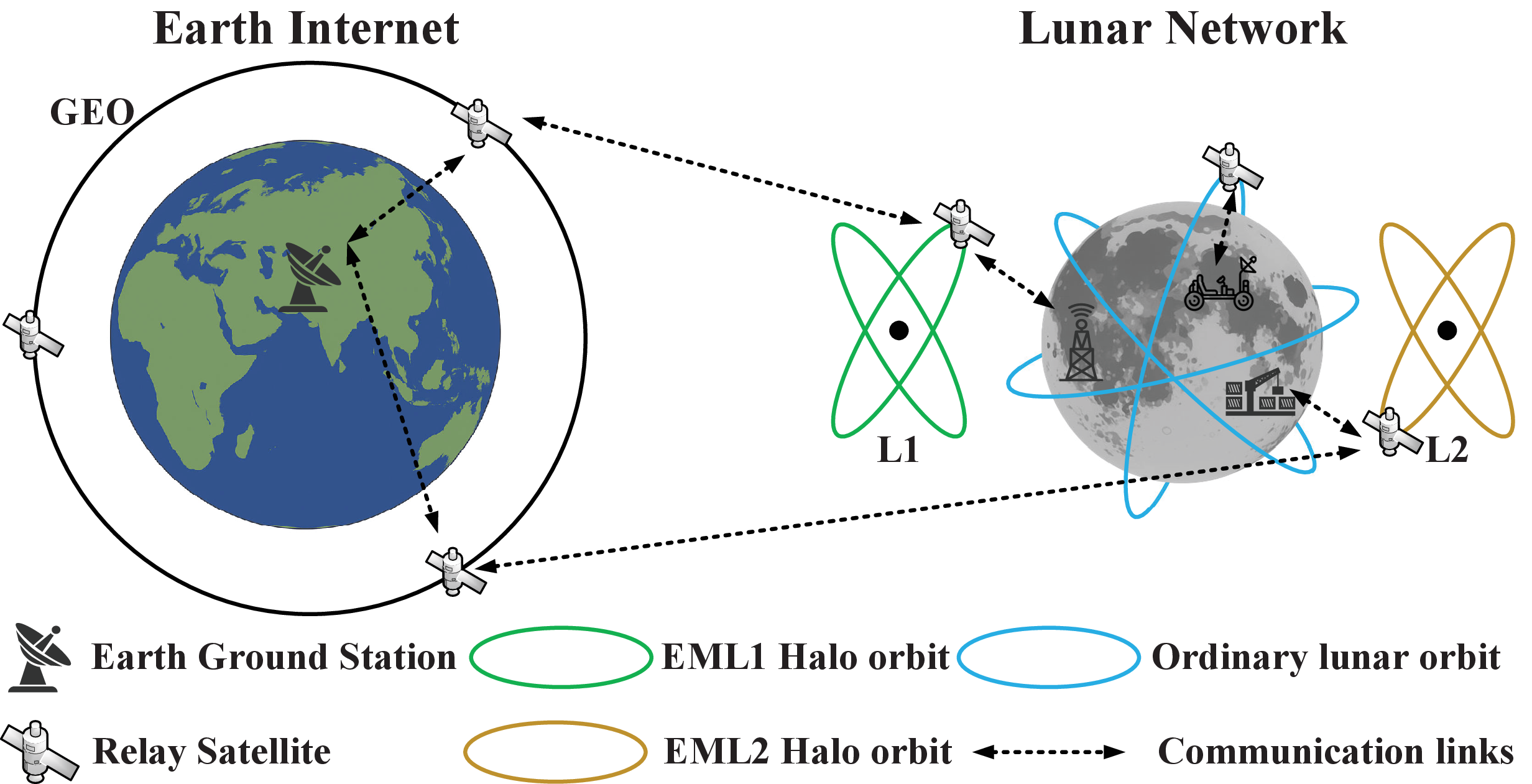}
    \caption{An Earth-Moon heterogeneous relay communication system.}
    \label{Earth_Moonrelay_system}
\end{figure}

\IEEEPARstart{I}{n} recent years, the global interest in deep-space and planetary exploration has continued to grow \cite{de2023roadmap}. Multiple space missions have been planned and implemented, such as NASA’s Artemis Program \cite{Artemis}, ESA’s Moonlight Initiative \cite{Moonlight}, China’s lunar exploration program \cite{ChangE5}, and its Mars exploration missions \cite{ChinaMars}. This growing interest has significantly stimulated research on interplanetary networks (IPNs) \cite{IPNs}. Consequently, the Interagency Operations Advisory Group (IOAG) has proposed communication architectures tailored to the Moon \cite{ioagmoon} and Mars \cite{ioagmars}, while NASA’s Space Communication and Navigation (SCaN) program has introduced the LunaNet \cite{Lunanet} to support long-term lunar activities. As an example of IPN, Fig. \ref{Earth_Moonrelay_system} illustrates an Earth–Moon heterogeneous relay communication system \cite{Joint}.

At present, the scale of deployed IPNs remains limited. For instance, the network supporting China’s Chang’e-5 mission comprised fewer than ten nodes \cite{ChangE5}. However, the communication requirements of deep-space missions are gradually expanding beyond scientific data return toward increasingly diverse applications, such as crewed lunar expeditions \cite{lunarexpeditions}, lunar permanent base operations and maintenance \cite{lunarbases}, and prospective Mars exploration and colonization \cite{gibney2022asteroids}. These applications will generate more complex traffic types, including real-time communication, navigation and positioning, and remote device control. Thus, the scale and service demands of IPNs are expected to grow rapidly in the near future \cite{Futurespacenetworks}.

Compared with terrestrial cellular and Earth-orbiting satellite networks, IPNs exhibit several unique challenges. First, the vast distances and relative motion between celestial bodies lead to extremely long and variable propagation delays. For instance, the round-trip time (RTT) between Earth and Mars can range from $8.5$ to $40$ minutes \cite{TCPinDSCN}. Second, deep-space links are subject to harsh channel conditions, characterized by high bit error rates (BER) and frequent packet losses. Finally, link interruptions, caused by factors such as planetary occlusion or severe electromagnetic environments, can last from minutes to days \cite{linkinterruptions}. These factors fundamentally undermine the performance of conventional Internet transport protocols.

To address these challenges, several approaches have been explored. The Delay/Disruption Tolerant Networking (DTN) architecture \cite{DTN}, based on a store-and-forward paradigm, is designed for deep-space communication. However, as IPNs expand and service demands diversify, the point-to-point model of DTN exhibits limited scalability and flexibility. Moreover, interworking with Internet Protocol (IP) networks requires protocol translation, which impedes the direct reuse of terrestrial networking technologies. To overcome this, the Taking IP To Other Planets (TIPTOP) working group advocates deploying and optimizing the IP stack in deep-space links \cite{IPArchitecture}. While this approach promises seamless integration, standard transport protocols such as Transmission Control Protocol (TCP) \cite{TCP} and Quick UDP Internet Connections (QUIC) \cite{rfc9000} suffer severe performance degradation because their mechanisms, like rapid loss recovery and congestion control, rely on low-latency acknowledgments (ACKs) and thus are unsuitable for the unique conditions of IPNs.

Performance Enhancing Proxies (PEPs) \cite{PEP} are a common solution for improving transport layer performance, widely employed in Low Earth Orbit (LEO) and Geostationary Earth Orbit (GEO) satellite networks. However, existing PEPs like PEPsal \cite{PEPsal} employ optimization techniques tailored to terrestrial and Earth-orbit environments and cannot accommodate the extreme propagation delays and frequent link interruptions of deep-space links. Furthermore, the widespread adoption of end-to-end encrypted protocols like QUIC is rendering traditional PEPs ineffective, as encrypting packet headers denies the inspection and manipulation upon which they rely.

To address these challenges, this paper proposes a transport-acceleration strategy for IPNs to improve the bandwidth utilization of TCP and QUIC while reducing end-to-end delivery delay. The strategy consists of a secure PEP architecture and a transport policy optimized for IPNs. Its key feature is the ability to perform connection splitting and performance enhancement for encrypted connections while preserving application-layer end-to-end security. The key contributions are as follows:

\begin{enumerate}
\item We propose a Non-Transparent Secure Proxy (NTSP) architecture. Through connection splitting and an independent application-layer data encryption mechanism, this architecture decomposes end-to-end encrypted connections to enable independent optimization of heterogeneous link segments for both TCP and QUIC protocols while preserving end-to-end application security.

\item We design an IPN-aware transport strategy to enhance the performance of the split TCP and QUIC connections. This strategy synergistically integrates two key components: 1) a rate-based congestion control algorithm (CCA) that leverages the pre-scheduled nature of deep-space links to ensure stable and efficient bandwidth utilization, and 2) an adaptive packet-level forward error correction (FEC) scheme, namely streaming codes (SC) \cite{DecodingCost}, to achieve low-latency loss recovery without retransmissions,  thereby decoupling loss recovery from congestion control.

\item We propose a theoretically-grounded backpressure flow control and buffer management mechanism. By modeling the proxy as a queuing system, we derive an analytical formula for the optimal buffer size. This mechanism effectively mitigates the rate mismatch between high-speed access links and low-speed deep-space links, preventing bufferbloat while balancing the trade-off between bandwidth utilization and queuing delay.

\item We implement and validate the proposed strategy through a prototype system named PEPspace. Extensive simulations are conducted in a representative Earth–Moon network. In a continuous connectivity scenario, the results demonstrate this strategy achieves near-capacity, stable goodput, significantly outperforming existing TCP/QUIC variants and open-source PEP solutions. Crucially, the proposed strategy avoids the severe link congestion and buffer bloat commonly observed in other high-performance schemes. In intermittent connectivity scenarios, this strategy effectively copes with link interruptions and achieves higher goodput than BP/LTP.
\end{enumerate}

Leveraging those designs, we further discuss a path towards a unified IP/DTN architecture, presenting a proof-of-concept to demonstrate its feasibility and identify key research challenges. The remainder of this paper is organized as follows. Section II reviews the related work. Section III presents our proposed secure acceleration strategy in detail. Section IV details our implementation and evaluates its performance. Section V explores the integration of IP and DTN architectures, outlining key challenges. Finally, Section VI concludes the paper and discusses future research directions. 

\section{Related Work}

\subsection{DTN Architecture}
DTN is designed to address intermittent connectivity and long propagation delays \cite{DTN}. The architecture is primarily composed of the Bundle Protocol (BP) \cite{BPv7} and the Licklider Transmission Protocol (LTP) \cite{LTP}. Its core mechanism is the construction of an overlay network through BP, which replaces the traditional end-to-end communication paradigm with a hop-by-hop forwarding model. The basic data unit, the bundle, follows a store-and-forward paradigm that allows persistent storage at intermediate nodes until forwarding opportunities arise, thereby providing tolerance to link disruptions. BP itself does not perform data transport but instead relies on Convergence-Layer Protocols (CLPs), such as LTP, TCP, or QUIC \cite{QUICL}, to achieve hop-by-hop delivery. Among these, LTP is the most commonly adopted CLP in IPNs \cite{LTP}. It forgoes congestion control, instead transmitting at a preconfigured rate. LTP encapsulates bundles into blocks and can operate in either reliable or unreliable modes. For reliable modes, an Automatic Repeat reQuest (ARQ) mechanism is used, with all retransmissions deferred until the end of a block transmission to reduce reverse-channel overhead.

Although extensive research has investigated the design and performance of BP and LTP \cite{HRBPRDS,Semantic-Aware,ASCLBP}, DTNs still face fundamental challenges in scalability and interoperability. Their overlay nature introduces substantial interoperability complexity. For example, in the lunar communication architecture proposed by IOAG \cite{ioagmoon}, IP-based Wi-Fi/5G networks are employed on the lunar surface, whereas DTNs are used over Earth–Moon links, necessitating intricate protocol translation for cross-domain communication. Such translation impedes the direct reuse of mature terrestrial networking applications. Consequently, while DTNs offer strong disruption tolerance, they sacrifice architectural integration with the mainstream IP ecosystem, resulting in a pronounced architectural mismatch for future, more interactive space missions.

\subsection{End-to-End Transport-Layer protocol Enhancements}
To address these challenges, the TIPTOP working group has advocated the direct deployment of an optimized IP protocol stack in deep-space links. By incorporating store-and-forward mechanisms at the IP layer, link disruptions can be mitigated effectively \cite{SRv6}. Nevertheless, extreme delays and high loss rates still severely impair transport-layer performance. Specifically, TCP struggles due to its default CCA, Cubic, which suffers from slow congestion window (CWND) growth in long propagation delays and frequent reductions under high packet loss, leading to poor bandwidth utilization. Building on this direction, a prominent line of research focuses on optimizing the CCAs. For instance, TCP Hybla \cite{Hybla} increases CWND by using a shorter reference RTT rather than the measured RTT, thereby improving goodput over GEO satellite links. Google’s BBR \cite{BBR} estimates the bandwidth–delay product (BDP) to determine CWND and achieves higher goodput than Cubic under random losses. However, in deep-space links, extremely long delays cause feedback signals to lag significantly behind link conditions, which constrains performance. Reinforcement learning–based schemes, such as OAC-TCPCC \cite{OAC-TCPCC}, have been proposed to address long delays and high loss, but they do not account for intermittent link disruptions. Furthermore, TCP relies on retransmissions for loss recovery, which introduces high Head-of-Line (HoL) blocking delays, as each recovery requires at least one RTT. This forces subsequent packets to wait in the receiver buffer, further degrading application-layer goodput. Although TIPTOP recommends QUIC \cite{rfc9000} as an alternative \cite{quic-profile}, QUIC inherits congestion control and loss recovery mechanisms similar to TCP, thereby facing comparable performance limitations in IPNs. 

\subsection{PEPs Overview}
PEPs offer an in-network solution by splitting connections. Their performance gain depends on the substitute transport mechanism used. However, current PEPs, such as PEPsal \cite{PEPsal}, design employ terrestrial-optimized CCAs that are ineffective under extreme delays and high loss rates characteristic of deep-space links. Moreover, with the growing adoption of encrypted protocols such as QUIC, traditional transparent PEPs have become ineffective, as they cannot parse encrypted protocol fields. This has led to explicit proxy designs. The Multiplexed Application Substrate over QUIC Encryption (MASQUE) framework \cite{MASQUE1,MASQUE2} allows clients to relay traffic through QUIC tunnels; however, proxies cannot access encrypted QUIC state information, and traffic is simply forwarded as opaque packets. Secure Multipath-Aware QUIC (SMAQ) \cite{SMAQ} redirects flows to proxies through enhanced connection migration. Yet, since QUIC connection migration requires completion of a $1$-RTT handshake, this approach introduces unacceptable overhead in IPNs. Therefore, in IPNs, effectively proxying encrypted protocols while preserving end-to-end security still requires the design of a dedicated proxy architecture.

\subsection{FEC for Loss Recovery}
Given that even a single retransmission over deep-space links may introduce delays of minutes to hours, FEC has been widely recognized as an effective alternative to reduce retransmission overhead. Multiple FEC schemes have been integrated into different protocol architectures. For instance, in DTN, the Erasure Coding Link Service Adapter (ECLSA) introduces Low-Density Parity-Check (LDPC) codes into LTP, showing promising results \cite{PLDTN}. In QUIC, QUIRL \cite{QUIRL} integrates coding schemes such as Reed-Solomon (RS) and window-based Random Linear Codes (RLC), achieving lower latency than standard QUIC under packet loss. In the PEP context, the open-source proxy \textit{Kcptun} \cite{KCP} implements RS-based FEC over a KCP protocol. 

However, these block-based coding schemes suffer from decoding delays that grow with block size, and residual losses still require retransmissions. In contrast, SC operates on a sliding window of packets, enabling recovery as soon as sufficient repair packets arrive, and has been shown to substantially reduce in-order decoding delay under fixed redundancy rates \cite{StreamingCoding}. For example, PEPesc \cite{PEPesc} is the first TCP proxy that applies SC for non-terrestrial networks (NTNs), whose prototype implementation\footnote{https://github.com/yeliqseu/pepesc.} has shown to be capable of achieving high and smooth goodput over GEO links

The review reveals a significant research gap: no existing solution simultaneously addresses the three major challenges of encrypted protocol optimization, extreme delay tolerance, and low-latency loss recovery. DTNs sacrifice interoperability, end-to-end optimizations are limited by RTT, and traditional PEPs lack deep-space-specific enhancements. Even advanced PEPs such as PEPesc exhibit slow CWND growth in IPNs and do not support proxy-based optimization for encrypted protocols such as QUIC. This motivates our work to design and validate a distributed acceleration strategy that integrates connection splitting, adaptive FEC, and deep-space-aware congestion control, while maintaining full compatibility with Internet encrypted protocols.





\section{A Secure Acceleration Strategy for Interplanetary Transport}

\subsection{Architecture Overview}
\begin{figure}[h]
    \centering
    \includegraphics[width=0.485\textwidth]{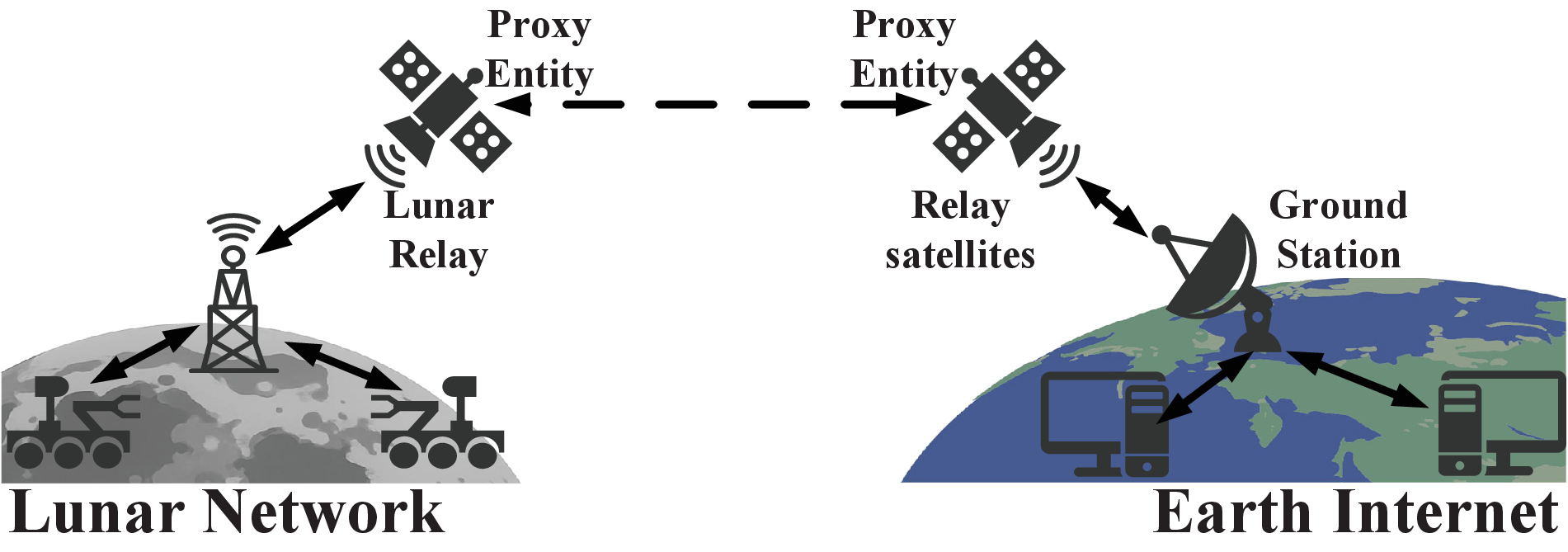}
    \caption{Architectural overview of the proposed secure acceleration strategy. Proxy entities are deployed at relay satellites to decompose an end-to-end connection into three independently optimized segments.}
    \label{system-modle02down}
\end{figure}

This section elaborates on the proposed secure acceleration strategy for IPNs. The key idea is connection splitting mechanism: PEPs are deployed at relay satellites to partition an end-to-end transport connection into independent sub-connections. As illustrated in Fig. \ref{system-modle02down}, in a representative Earth-Moon network, proxies on Earth-side and Moon-side relay satellites terminate access connections at segment boundaries and establish an optimized transport tunnel across the interplanetary segment. For clarity, we focus on the downstream traffic from Earth to the Moon in the following subsections; however, the architecture is fully symmetric and supports bidirectional acceleration.

The proposed strategy rests on three core components. Section~III-A presents the NTSP architecture that provides a secure foundation for splitting encrypted connections. Section~III-B describes the IPN-aware transport policy, which combines rate control with low-latency loss recovery. Section~III-C details the backpressure flow-control and buffer-management mechanisms that ensure system stability and prevent buffer bloat. 

QUIC is selected as the underlying transport protocol between proxy entities. Its native stream multiplexing capabilities allow traffic from multiple access connections to be efficiently aggregated and transported, while its user-space implementation provides the flexibility required to deploy our custom IPN-aware transport policy. Importantly, one of NTSP’s key principles is transport-layer modularity, which allows the underlying transport protocol to be replaced or extended for other protocols as needed.

\subsection{The NTSP Architecture}
The first pillar of our proposed strategy is the NTSP architecture designed specifically for IPNs. At its core, NTSP deploys proxy entities along deep-space links and employs a connection-splitting mechanism to decompose an end-to-end transport path into multiple heterogeneous segments. These proxies intercept and terminate transport connections at segment boundaries to perform protocol adaptation and apply deep-space-specific optimizations. Because QUIC packets are fully encrypted, NTSP requires endpoints to trust proxy nodes at the protocol-operation level for connection splitting. It should be emphasized that this trust is constrained: proxies are prohibited from accessing application-layer data carried in QUIC STREAM frames. Moreover, unlike network-layer automatic interception mechanisms, NTSP is invoked only upon explicit endpoint selection, thereby mitigating the ossification caused by traditional PEPs. After an NTSP connection is established, the endpoint further authorizes the selected proxy node to make optimal forwarding decisions at the next hop—specifically, whether to relay traffic to a more suitable node or to forward it directly to the destination host. This is because a proxy node may be better positioned than the endpoints to select the optimal path for a given flow.

We use QUIC proxying as an example to illustrate the NTSP connection-establishment procedure and its associated overhead. We assume that historical connections among proxy nodes have been established in advance, enabling $0-$RTT connection establishment. Additionally, the client has already discovered the proxies’ addresses, and that each proxy knows its optimal forwarding path.

\begin{figure}[t]
    \centering
    \includegraphics[width=0.485\textwidth]{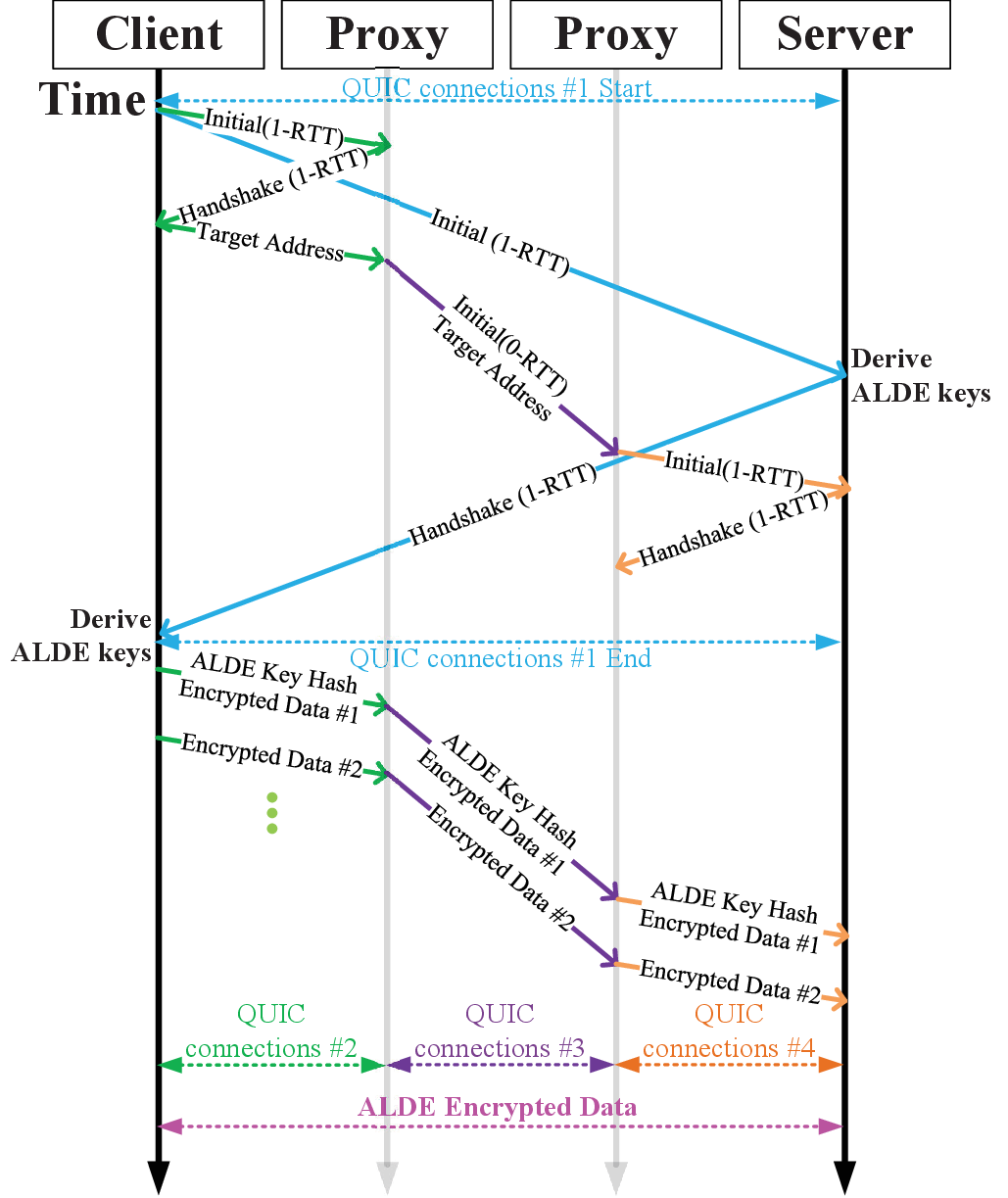}
    \caption{NTSP $1$-RTT Establishment Procedure.}
    \label{1-RTT}
\end{figure}

\subsubsection{\textbf{Connection Setup}} \label{Connection_Setup}
The NTSP connection establishment procedure is illustrated in Fig. \ref{1-RTT}. First, the client and server perform a $1-$RTT handshake (Fig. \ref{1-RTT}, blue arrows) to establish a key-derived connection, hereafter referred to as Connection $\#1$. During the handshake, both endpoints need to indicate support for NTSP via \texttt{NTSP} transport parameters in their initial packets. Subsequently, the server and client each derive the Application Layer Data Encryption (ALDE) key, thereby constructing an additional end‑to‑end encryption layer on top of QUIC’s native Transport Layer Security (TLS) protection. This ALDE key is derived from the TLS $1.3$ handshake, maintained exclusively by each endpoint and is never transmitted. The specifics of the encryption mechanism will be detailed in section \ref{Application_Layer_Data_Encryption}.

Concurrently with the initiation of Connection $\#1$, the client initiates a second $1-$RTT handshake with the selected proxy node (Fig. \ref{1-RTT}, green arrows), denoted Connection $\#2$. Over Connection $\#2$, the address of the target server is sent to the proxy by the client. Upon receipt, connectivity to the next‑hop peer is evaluated by the proxy. If the next hop is another proxy and no connection exists, a $0-$RTT handshake (Fig. \ref{1-RTT}, purple arrows) is initiated by the proxy, establishing Connection $\#3$ and forwarding the target address. Otherwise, an existing QUIC connection is reused by opening a new stream. If the proxy's next hop is the target server, a $1-$RTT handshake (Fig. \ref{1-RTT}, orange arrows) is performed to establish Connection $\#4$. A critical design distinction from terrestrial proxies is that this process requires no explicit path validation feedback to the client. This departure is fundamental to the IPN context, where persistent end-to-end connectivity is not guaranteed. Each proxy node is designed with store-and-forward capabilities, assuming full responsibility for eventual data delivery to the next hop. This procedure similarly applies to opening new streams after the NTSP connection is established. In this scenario, the proxy maintains a mapping between the streams in Connection $\#2$ and Connection $\#4$. For each incoming stream, the proxy creates a corresponding proxy stream, and directs it to a specific connection through the mapping relationship. By this mapping relationship, NTSP preserves QUIC’s stream multiplexing benefits and avoids HoL blocking when proxying multiple connections.

Upon establishment of Connection $\#1$, the ALDE key is derived, permitting closure of Connection $\#1$. Thereafter, three independent connection segments remain: client‑to‑proxy (Connection $\#2$), proxy‑to‑proxy (Connection $\#3$), and proxy‑to‑server (Connection $\#4$). The end‑to‑end connection is thus decomposed into three autonomous, forwardable sub‑connections (green, purple, and orange layers in Fig. \ref{1-RTT}). In this architecture, the server‑side proxy acts as the virtual client and the client‑side proxy acts as the virtual server, jointly reconstructing the original connection. Each sub‑connection’s transport security is upheld by QUIC, while end‑to‑end confidentiality of application data is maintained by the ALDE encryption layer (magenta layer in Fig. \ref{1-RTT}).

\begin{figure}[t]
    \centering
    \includegraphics[width=0.485\textwidth]{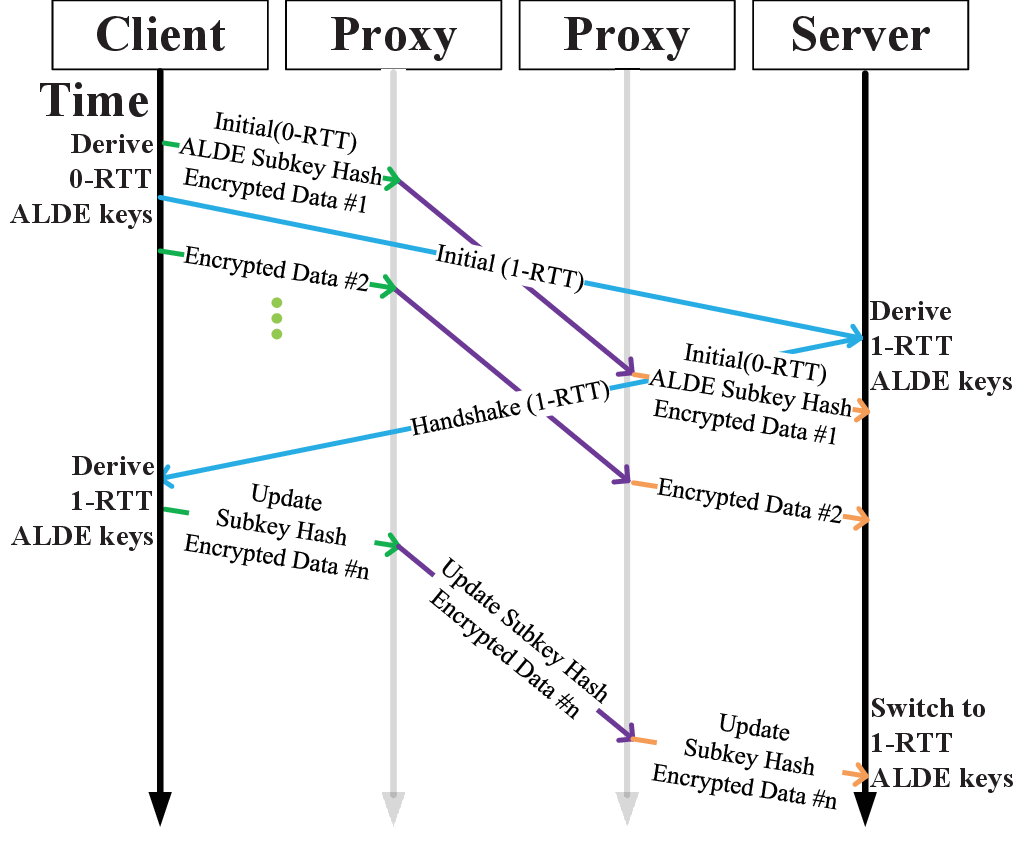}
    \caption{NTSP $0$-RTT Establishment Procedure.}
    \label{0-RTT}
\end{figure}

NTSP also supports $0-$RTT connection establishment, as shown in Fig. \ref{0-RTT}. In this variant, the client first performs a $0-$RTT handshake with the proxy and directly sends encrypted data. The ALDE key used is derived from the session resumption ticket of historical Connection $\#1$. Meanwhile, the client still needs to complete a $1-$RTT handshake with the server to establish a fresh ALDE key. Upon its derivation, the new ALDE key must be adopted immediately to ensure forward secrecy and replay protection.

\subsubsection{\textbf{Application Layer Data Encryption Mechanism}}\label{Application_Layer_Data_Encryption}

\begin{figure}[t]
    \centering
    \includegraphics[width=0.485\textwidth]{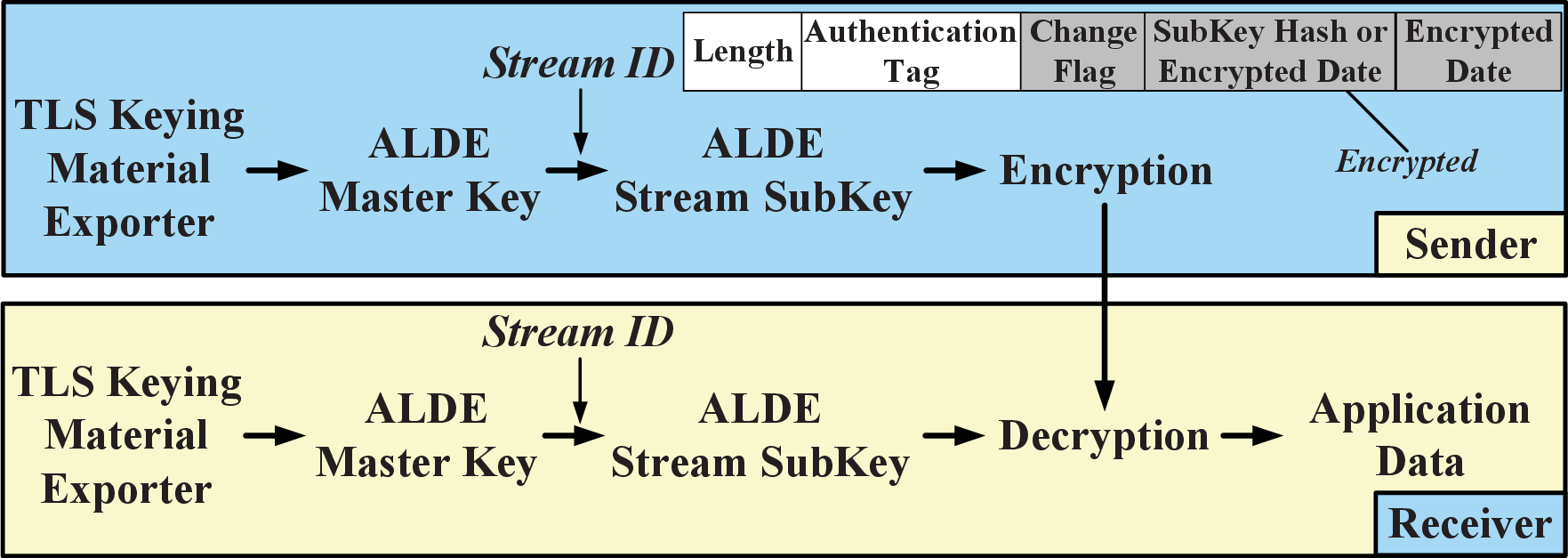}
    \caption{NTSP key derivation process.}
    \label{Key_export_process}
\end{figure}

QUIC employs a unified TLS session key to protect both application data and control signaling. Consequently, native end‑to‑end encryption is broken when the connection is split. To ensure application data security, an independent encryption layer—termed ALDE—is introduced. Through ALDE, proxy nodes are restricted to processing only QUIC protocol headers and control frames, while application‑layer payloads remain encrypted and are immune to decryption, tampering, or replay. 

Fig. \ref{Key_export_process} illustrates NTSP’s key derivation process. The ALDE master key is exported via the TLS $1.3$ Keying Material Exporter \cite{RFC5705} from the TLS‑derived keying material of a QUIC session. This exporter permits secure derivation of auxiliary keys after handshake completion without the overhead of an additional TLS handshake, and without compromising the primary TLS channel. The key derivation process of ALDE also provides forward secrecy: each generation of derived keys depends solely on its predecessor, such that a newly derived key cannot recover any earlier key. During transmission, a subkey is derived for each QUIC stream from the ALDE master key to encrypt the data on that stream. Because QUIC streams are bidirectionally symmetric, both traffic directions on the same stream are protected by a single subkey.

Advanced Encryption Standard–Galois/Counter Mode (AES‑GCM) is employed for encryption, with each invocation using a unique nonce derived from a per‑stream, monotonically increasing sequence number. The receiver accepts only ciphertext chunks whose nonces are fresh and in proper sequence, thereby preventing replay attacks. The format of an ALDE–encrypted block is depicted in Fig. \ref{Key_export_process} and comprises four fields in its header:
\begin{enumerate}
\item \textbf{Length} ($2$ bytes): the length of the encrypted payload;
\item \textbf{Authentication Tag} ($16$ bytes): for integrity and authenticity verification;
\item \textbf{ChangeFlag} ($1$ byte): indicates triggering key update;
\item \textbf{Encrypted Data}: the ciphertext of the application payload. When ChangeFlag is $1$, the initial portion of Encrypted Data carries a $32‑$byte hash of the new subkey, enabling the receiver to synchronize its decryption context.
\end{enumerate}

Because the ALDE key-derived connection (Connection $\#1$) is independent of the data‑forwarding connections (Connections $\#2$–$\#4$), each stream’s initial subkey is identified by including its $32‑$byte hash in the stream header. As the hash function is preimage‑resistant, identifiers can be transmitted in clear text. Upon receipt, the corresponding subkey is located in the server’s local cache based on the hash, and subsequent ciphertext blocks are decrypted.

Subkey updates are performed on a per-stream basis at the sender’s discretion. When a key refresh is required, the next subkey is derived from the current one using the TLS HMAC-Based Key Derivation Function (HKDF). A block with its ChangeFlag set to $1$ carries the hash of the new subkey. This block is encrypted under the old subkey, allowing the receiver to update its local key mapping. In $0-$RTT mode, the initial ALDE master key is exported from the session resumption  ticket of the historical Connection $\#1$. Following completion of a fresh end to end $1-$RTT handshake, the client must transitions to a new ALDE master key derived from the updated session keys to preserve forward secrecy and mitigating replay threats.

\subsubsection{\textbf{Connection Overhead}}\label{Connection_Overhead}
In NTSP, connection‑handshake overhead is governed by the one‑way propagation delays $D_0$ (client to server), $D_1$ (client to proxy), $D_2$ (proxy to proxy), and $D_3$ (proxy to server). In the standard QUIC $1$‑RTT handshake, application data can be sent immediately upon receipt of the ServerHello, incurring approximately $1\times$RTT of handshake overhead (i.e., $2D_0$). In NTSP, completion of Connection $\#1$ must be awaited by the client to obtain the ALDE master key before encrypted application data can be transmitted, resulting in a fixed delay of $2D_0$. Simultaneously, the handshake of Connections $\#2$ and $\#4$ incur delays of $2D_1$ and $2D_3$, respectively, whereas Connection $\#3$ incurs an overhead of $D_2$. As these segments can be established in parallel, the total handshake latency overhead can be approximated as: 
\begin{equation}
max(2D_0, 2D_1+D_2+2D_3).
\end{equation}
Thus, if the direct path is longer, the end‑to‑end delay ($2D_0$) governs; otherwise, the proxy chain delay ($2D_1 + D_2 + 2D_3$) prevails. In IPNs, where the proxy is typically located on the direct path (i.e., $D_0 = D_1 + D_2 + D_3$), NTSP’s handshake latency remains equivalent to one end-to-end RTT, which is the same as that of standard QUIC. Moreover, if a historical NTSP session exists, NTSP can leverage $0-$RTT handshake to avoid extra round trips, analogous to QUIC’s native $0-$RTT.

Additionally, as mentioned earlier, each ALDE‑encrypted block contains a fixed header overhead comprising a $2‑$byte length field, a $16‑$byte authentication tag, and a $1‑$byte ChangeFlag identifier, for a minimum of $19$ bytes. Although ALDE‑encrypted data length is variable, ALDE stipulates that the maximum length of a single encrypted block equals that of TLS ($2^{14}$ bytes) to balance security and implementation compatibility. Therefore, ALDE encryption incurs a minimum bandwidth overhead of approximately $0.11\%$.

\begin{table}[t]
	\centering
	\caption{COMPARISON OF EXPLICIT PROXY APPROACHES.}
	\begin{tabular}{>{\centering\arraybackslash}m{1.5cm} | c| c | c } 
		\toprule
		\textbf{Scheme} & \textbf{MASQUE} & \textbf{SMAQ} & \textbf{NTSP} \\ 
        \midrule
		\textbf{Handshake Overhead} & $2D_{0}+2D_{1}$ & $> 4D_{0}$ & $2D_{0}$ \\ \hline
		\textbf{Bandwidth Overhead} & $0\%$ & $0.13\%$ & $0.11\%$ \\ \hline
		\textbf{Connection Splitting} & No & Support & Support \\ \hline
		\textbf{End-to-End Security} & Full & Application layer & Application layer \\ \hline
		\textbf{0-RTT Handshake} & No & No & Support \\ 
        \bottomrule
	\end{tabular}
	\label{tab2}
\end{table}

Table~\ref{tab2} compares MASQUE, SMAQ, and NTSP in scenarios where the proxy resides on the direct path. MASQUE is deployed on a client-side proxy, while SMAQ, similar to NTSP, utilizes a distributed deployment with proxies at both ends of the connection. The results indicate that NTSP is the only proxy solution supporting 0-RTT mode while also achieving lowest handshake overhead. By employing AES-GCM instead of the TLS record protocol used in SMAQ, NTSP avoids protocol complexity and attains slightly reduced bandwidth overhead. Furthermore, unlike MASQUE's opaque forwarding, NTSP's connection splitting enables fine-grained optimization of each transport segment, crucial for heterogeneous deep-space links with vastly different characteristics.

\subsubsection{\textbf{Security Considerations}}\label{Security_Considerations}
NTSP's security relies on end-to-end application-layer encryption and a defined trust model for intermediary proxies. Two attacker classes are considered. The first class comprises honest-but-curious proxies that faithfully follow protocol specifications but attempt to glean additional information through eavesdropping and traffic analysis. The second class comprises malicious proxies that may engage in active attacks such as payload tampering, replay, dropping, or delaying of packets. NTSP's cryptography ensures that application data confidentiality and integrity are preserved even in the presence of malicious proxies; however, availability protection remains dependent on the trust constraints imposed on proxies.

NTSP's core security mechanism is the secure derivation and isolation of ALDE keys. The ALDE master key is derived within the native QUIC end-to-end handshake; this operation is performed independently of any intermediary proxy, ensuring that proxies cannot intercept or learn the master key. For inter-stream key association, each stream is associated with a key identifier formed by the 32-byte hash of its ALDE subkey; the identifier is transmitted in the clear to enable correct key binding across streams. The hash’s one-way property prevents recovery of the underlying subkey from the identifier. Servers perform efficient identifier validation via a local cache, which mitigates risks of key forgery and hash collisions.

Architecturally, an end-to-end connection is partitioned into three independent sub-connections. Proxy nodes handle transport-layer context and QUIC signaling, whereas application payloads remain encrypted end-to-end under ALDE at all times. AES-GCM authentication tags are used to provide tamper resistance; any payload altered in transit will fail authentication at the receiver and be discarded. Per-stream monotonically increasing nonces are used to prevent replay attacks. Consequently, even in the presence of malicious proxies, NTSP provides end-to-end confidentiality, integrity, and replay protection for application data.

Several limitations remain. NTSP does not conceal metadata; therefore, traffic analysis based on packet sizes and timing remains feasible. NTSP also does not address transport-layer interference: malicious proxies may still degrade availability by dropping or delaying encrypted payloads or by manipulating transport-layer QUIC signaling within the segments they control. Although such actions do not compromise data confidentiality, they may cause connection interruptions, performance degradation, or anomalous application behavior.

Accordingly, availability guarantees in NTSP require management-level trust in deployed proxies. Specifically, this trust model requires that proxies faithfully forward packets and adhere to the protocol's transport-layer signaling semantics within their governed segments. The design of NTSP is leveraging cryptography to downgrade the trust required of proxies: from guaranteeing data security to ensuring only service availability. This design is particularly suited for IPNs, which are collaborative yet mutually distrustful environments. In these settings, NTSP enables different space agencies to share communication links based on operational trust, while cryptographically guaranteeing the confidentiality and integrity of their mission data against each other.

\begin{figure*}[t]
    \centering
    \includegraphics[width=0.98\textwidth]{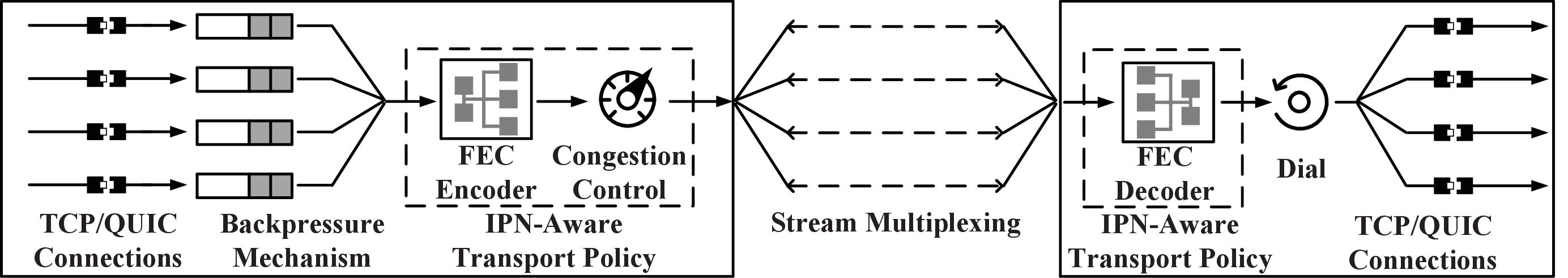}
    \caption{Functional block diagram of a proxy entity.}
    \label{system-modle02up}
\end{figure*}

\subsection{An IPN-Aware Transport Strategy}
Following the provision of secure connection splitting capabilities by the NTSP architecture, the second core component of our strategy is an IPN-aware transport strategy, which can then be deployed over the deep-space segment to enhance end-to-end performance. Unlike the passive probing mechanisms commonly used in terrestrial networks, this strategy exploits the pre-scheduled characteristics of IPN links and adopts a proactive transmission approach that alleviates dependence on feedback signals under long propagation delays. Fig.~\ref{system-modle02up} illustrates the functional architecture of this policy within a proxy entity. It consists of two core components: First, a rate-based congestion control algorithm is employed to achieve high bandwidth utilization. Second, an adaptive FEC mechanism provides low-latency loss recovery without retransmissions.

\subsubsection{\textbf{Rate-based CCA between PEPspace Entities}}
Conventional CCAs that rely on ACK feedback perform poorly in IPNs because feedback signals lag far behind actual link conditions. In contrast to terrestrial networks, key parameters such as link capacity and propagation delay in IPNs are typically pre-scheduled and predictable. Moreover, because packet loss is fully recovered through FEC, the recovery latency is decoupled from the RTT, allowing congestion control to be simplified to pure rate control. 

Consequently, our strategy abandons complex probing procedures and adopts a straightforward CCA based on a fixed-bandwidth model. This algorithm calculates the link BDP using the pre-scheduled link capacity and the measured minimum RTT, and directly sets the CWND accordingly. Such a planned-rate approach enables the data flow to quickly reach and stably maintain near-capacity goodput, fundamentally avoiding the slow start-up and severe oscillations that plague traditional CCAs in long-delay environments.

\subsubsection{\textbf{FEC coding/decoding mechanism}}
To mitigate the significant HoL latency caused by long propagation delay, our strategy adopts SC to achieve low-latency loss recovery without retransmissions. All QUIC frames that require FEC protection are encapsulated into a newly defined SOURCE frame, denoted $S_i$, where $i$ denotes the sequence index (ID). Denote $i_{seq}$ as the ID of the latest transmitted SOURCE frame. Initially, $i_{seq}$ is set to -1. After each SOURCE frame is transmitted, $i_{seq}$ is incremented by $1$. The REPAIR frame is constructed as a weighted linear combination of all unacknowledged SOURCE frames over the finite field $\mathbb{F}_{2^m}$. It is generated as $R_{k} = \sum_{i=w_s}^{i_{seq}} a_{k,i} S_{i} ,$ where $k = 0, 1, \dots$ indicates the REPAIR frame ID. The coding coefficients $a_{k,i}$ are selected from $\mathbb{F}_{2^m}$ by a pseudo-random number generator (PRNG). With an agreed or pre-shared seed, $a_{k,i}$ can be regenerated without explicit transmission. The parameter $w_s$ denotes the lowest ID of all transmitted but unacknowledged SOURCE frames and is dynamically updated upon acknowledgments. The $[w_s, w_e]$, where $ w_e \equiv i_{seq} $, is defined as the Encoding Window (EW). Each REPAIR frame carries its EW bounds.

Upon a packet sending opportunity, the insertion of REPAIR frames is determined based on the status of the sending queue. Specifically, if the proxy's send buffer contains pending data, a REPAIR frame is inserted based on the estimated loss rate $p_e$ inferred from real-time QUIC feedback. The decision to send a REPAIR frame is made dynamically based on the current redundancy ratio, defined as $N_{r}/(N_{s}+N_{r})$. A new REPAIR frame is transmitted if this ratio is below a target threshold determined by $p_e$:
\begin{equation}
\label{eq:repair_condition}
\begin{matrix}N_{r}/(N_{s}+N_{r})< p_{e}+\delta
  &\delta \in (0,1),
\end{matrix}
\end{equation}
where $\delta$ is a small redundancy compensation factor that is inversely related to the decoding delay \cite{StreamingCoding}.

When the send buffer is empty but SOURCE frames are still unacknowledged, the mechanism actively counteracts tail loss. In this state, REPAIR frames continue to be transmitted as long as congestion control allows, until all SOURCE frames are acknowledged. These "tail-protection" REPAIR frames are sent during idle periods, thus not degrading the goodput of normal traffic. Note that REPAIR frames sent solely while waiting for acknowledgments are excluded from the $N_r$ count to prevent throttling of new REPAIR frame generation when fresh data arrives. 

At the receiver, the decoder tracks the latest in-order received SOURCE frame ID $i_{ord}$. When the next received frame is neither $S_{i_{ord}+1}$ nor a REPAIR frame covering $w_e \ge i_{ord}+1$, a loss is detected and the decoder is activated. The on-the-fly Gaussian elimination (OF-GE) algorithm \cite{OFGE} is then employed to recover lost SOURCE frames. Define $ \hat{w}_s \equiv i_{ord} + 1 $ and let $ \hat{w}_e $ denote the largest $ w_e $ among the EWs of the received REPAIR frames. The interval $[ \hat{w}_s, \hat{w}_e ]$ is referred to as the decoding window (DW) of the current decoder. The DW expands dynamically as more frames arrive. Once the number of linearly independent REPAIR frames equals the number of missing SOURCE frames within the DW, the OF-GE decoding succeeds, recovering all lost SOURCE frames. Note that although the decoder buffers a copy of the SOURCE frame for decoding, delivery of the SOURCE frame to proxy is never blocked. This design ensures that the common FEC mechanism across multiple proxy streams does not break QUIC’s stream-multiplexing semantics.

\subsection{Backpressure Flow Control and Buffer Management} \label{KOPT}
To ensure the stable operation of the entire acceleration strategy and prevent system breakdown caused by link rate mismatch, the third key pillar of our strategy is a theoretically grounded backpressure flow control and buffer management mechanism. The connection‑splitting mechanism fundamentally decouples the native congestion control feedback loop. Specifically, upon receiving data from the sender, proxy immediately issues a local acknowledgment to the sender, so called because the data has not arrived at the original destination yet. This prevents the sender from observing downstream conditions, particularly the inherent high latency and loss rates of deep‑space segments. Thus, the sender may overestimate available bandwidth and inflate its CWND, provoking data avalanches into proxy node and risking buffer exhaustion.

To mitigate this risk without altering the sender protocol, backpressure is applied via standard flow‑control mechanisms. Unlike the dynamically updated CWND, a flow‑control window advances only as the receiver consumes data. When proxy’s buffer reaches capacity, the sender is forced to stall until proxy consumes data from buffer, thereby constraining the send rate to match the proxy’s forwarding capacity.

However, flow‑control–based backpressure introduces additional queuing delay: packets must wait in the proxy buffer before dispatch. If the buffer size is set too small, the transmission rate may be insufficient to utilize the full path bandwidth; if the buffer size is too large, excessive queuing delay will degrade latency‑sensitive traffic. Thus, buffer size must be chosen to balance maximal bandwidth utilization against minimal queuing delay.

Classical guidelines recommend sizing the buffer to the end‑to‑end BDP. Yet, in a split‑connection architecture like NTSP, this rule is ambiguous and suboptimal: it is unclear whether to use the client–proxy or proxy–server segment’s bottleneck bandwidth and RTT, or some combination thereof. To quantify the trade‑off between bandwidth utilization and queuing delay, we develop an analytical model that incorporates input‑segment latency characteristics and output‑segment bandwidth constraints to determine the optimal proxy buffer size.

For closed‑form tractability and intuitive insight, we first assume per‑stream packet arrivals follow a Poisson process and service times (i.e., per‑packet forwarding delays) are exponentially distributed. Under these assumptions, each stream’s buffer can be modeled as an $M/M/1/K$ queue. Let the bandwidth and RTT of the proxy's ingress link be denoted by $B_{in}$ and $RTT_{in}$, respectively, and those of the egress link be $B_{out}$ and $RTT_{out}$. With $N$ concurrent streams, each maintaining a flow‑control window of size $K$, the per‑stream service rate
\begin{equation}
\mu =\frac{BW_{out}}{N},
\end{equation}
which is determined by the fair share of the proxy’s egress bandwidth $BW_{out}$. The arrival rate $\lambda$ is limited by the lesser of the ingress-segment bandwidth share and the flow-control window rate:
\begin{equation}
\lambda=min(\frac{BW_{in}}{N} ,\frac{K}{RTT_{in}}).
\end{equation}
Hence, the system load
\begin{equation} \label{RHO}
\rho=\frac{\lambda}{\mu}=min(\frac{BW_{in}}{BW_{out}},\frac{K*N}{BW_{out}*RTT_{in}}).
\end{equation}
Since $BW_{in} \gg BW_{out}$ in IPNs, the second term in equation (\ref{RHO}) governs system behavior. 

The steady‑state empty‑buffer probability $p_0$ and utilization $U=1-p_0$ follow standard $M/M/1/K$ results:
\begin{equation}
p_0=\left\{\begin{matrix} \frac{ 1-\rho}{ 1-\rho^{K+1}} ,
  &\rho \ne  1,\\ \frac{1}{K+1} ,
  &\rho = 1.
\end{matrix}\right.
\end{equation}

\begin{equation}
U=1-p_0=\left\{\begin{matrix} \rho \frac{ 1-\rho^{K}}{ 1-\rho^{K+1}} ,
  &\rho \ne  1,\\ \frac{K}{K+1} ,
  &\rho = 1.
\end{matrix}\right.
\end{equation}

Three operational regimes emerge:
\begin{enumerate}
  \item \textbf{Under‑buffered} ($K< \frac{BW_{out}*RTT_{in}}{N} \Rightarrow \rho<1$): Goodput is constrained by $K$, and $U\approx \rho$ grows nearly linearly with $K$, but the egress link remains underutilized.
  \item \textbf{Optimal point} ($K = \frac{BW_{out} \cdot RTT_{in}}{N} \Rightarrow \rho=1$): The system achieves balance, with $U = \frac{K}{K+1}$ approaching unity and a uniform distribution of queued packets. The mean system occupancy $L=\frac{K}{2}$. To determine the average sojourn time $W$, we apply Little's Law, $W = L/\lambda_{eff}$, where $\lambda_{eff}$ is the effective arrival rate. In steady state, the effective arrival rate equals the departure rate, which is the product of the service rate $\mu$ and the utilization $U$. Therefore, $\lambda_{eff} = \mu \cdot U = \mu \frac{K}{K+1}$. The average sojourn time is thus:
\begin{equation}
 W = \frac{L}{\lambda_{eff}} = \frac{K/2}{\mu \frac{K}{K+1}} = \frac{K+1}{2\mu}.
\end{equation}
This shows that at the optimal point, queuing delay is equivalent to half the time it takes to drain a full buffer.
\item \textbf{Over-buffered} ($K> \frac{BW_{out}*RTT_{in}}{N} \Rightarrow \rho>1 $): The buffer is excessive, the egress link saturates, and additional $K$ yields marginal utilization gains ($U\to1$), but queuing delay grows linearly as $W = \tfrac{K}{\mu}$.
\end{enumerate}

To maximize bandwidth utilization while preventing excessive queuing delay, operation at the critical load $\rho = 1$ is recommended, yielding the optimal buffer size

\begin{equation}\label{EQKPOT}
K_{opt} = \frac{BW_{out}*RTT_{in}}{N}.
\end{equation}

Recognizing that real‑world packet sizes approximate the maximum transmission unit and that forwarding delays are relatively constant, an $M/D/1/K$ model offers greater physical fidelity. Although its steady-state probability distribution differs from that of the $M/M/1/K$ model, the critical point, where the system attains maximum utilization and queue length begins to grow linearly, still occurs at a system load of $\rho \approx 1$. The following analysis thus serves to validate the $K_{opt}$ formula under these more realistic assumptions. This model assumes a deterministic service time of $1/\mu$. Since the service time is deterministic, the system is no longer a pure birth–death process and must be analyzed using the embedded Markov chain method. Following the classical results in \cite{XQueue}, the steady-state probability $\pi_n$, which corresponds to $p_{n}$ in $M/M/1/K$, is given by:

\begin{equation}\label{pin}
\pi_n 
= C\;\frac{\rho^n}{n!}\;
\sum_{k=0}^{K-n}\frac{(-\rho)^k}{k!}
\quad
(0\le n\le K),
\end{equation}
where $C$ is the normalization constant satisfying:
\begin{equation}\label{C-1}
C^{-1}
=\sum_{i=0}^K\frac{\rho^i}{i!}\sum_{k=0}^{K-i}\frac{(-\rho)^k}{k!}.
\end{equation}
The system's bandwidth utilization, $U = 1 - \pi_0$, is the same as that of the $M/M/1/K$ model. The average number of packets in the system, $L$, is calculated as:
\begin{equation}\label{L}
L = \sum_{n=1}^K n\,\pi_n.
\end{equation}
Applying Little's Law, the average sojourn time $W$ is then determined by:
\begin{equation}\label{W}
W = \frac{L}{\lambda\,(1-\pi_K)}.
\end{equation}
A numerical evaluation of Eq. (\ref{pin})-(\ref{W}) demonstrates that the optimal buffer size remains $K_{\text{opt}}$ for the M/D/1/K model.

Compared to the traditional end‑to‑end BDP heuristic, this formula—by explicitly combining proxy egress bandwidth and ingress RTT—captures the decoupling effects of the split‑connection architecture and provides precise guidance for PEP parameter configuration.

\section{IMPLEMENTATION AND PERFORMANCE EVALUATION}

\subsection{Prototype Implementation}
We have implemented our strategy in a prototype system named PEPspace. The prototype is developed in Go and is built upon the widely-used open-source \textit{quic-go} library. This foundation allows for flexible manipulation of transport-layer mechanisms in user space. 

The connection setup, key derivation, and ALDE mechanisms follow the procedures described in Section III-A. To preserve QUIC’s native stream multiplexing across the proxy, a dedicated proxy stream is established for each incoming access stream. A simple round-robin scheduler reads data from these access streams. While this scheduler ensures basic fairness, the architecture can readily accommodate more advanced algorithms (e.g., Weighted Fair Queuing) to support differentiated Quality of Service (QoS).

Data read from all streams is then processed by the unified IPN-aware transport policy. The adaptive SC is implemented over the finite field $\mathbb{F}_{2^{8}}$, with a redundancy compensation factor of $\delta=0.03$. A key feature of the implementation is that FEC encoding operates across all multiplexed streams: data from low-rate streams are protected by redundancy generated from high-rate streams, thereby enhancing overall reliability. The rate-based CCA sets the congestion window (CWND) according to the pre-configured bandwidth.

Finally, to ensure system stability, the access-side flow-control window for each incoming stream is configured based on the optimal buffer size $K_{opt}$ derived from the theoretical model in Section III-C, ensuring a stable trade-off between goodput and latency.

\subsection{Simulation setup}
\begin{figure}[h]
    \centering
    \includegraphics[width=0.485\textwidth]{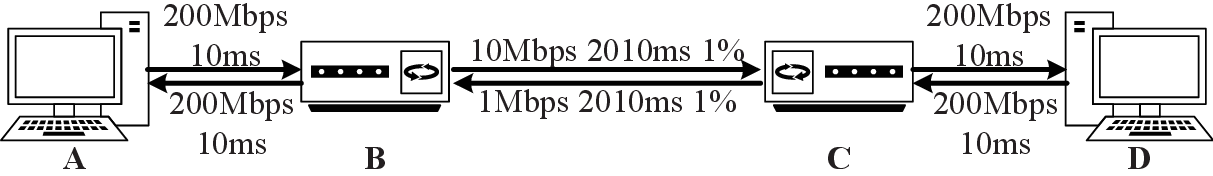}
    \caption{Network simulation topology, where the B-C link emulates the Earth-Moon interplanetary bottleneck.}
    \label{NetworkTopo}
\end{figure}

This section details the simulation methodology used to evaluate the performance of our proposed strategy. The evaluation is conducted using the Mininet emulator with a four-node linear topology, as illustrated in Fig. \ref{NetworkTopo}. The link between nodes B and C is configured to emulate the Earth–Moon interplanetary bottleneck. Considering the asymmetric bandwidth characteristics of Earth–Moon communications, the downlink capacity is set to $10$ Mbps and the uplink capacity to $1$ Mbps, with a one-way propagation delay of $2.01$ seconds. Various packet loss conditions are also applied. Since pure interplanetary links are not affected by stochastic channel fading factors such as atmospheric effects \cite{ACTIS}, their packet loss events can be regarded as independent \cite{Joint}. More sophisticated loss models require precise deep-space channel characterization and impose higher demands on FEC coding and control mechanisms; such extensions are left for future work. To eliminate the impact of non-interplanetary links, the access links (AB and CD) are configured with $200$ Mbps bandwidth, $10$ ms latency, and no packet loss.

Two representative interplanetary scenarios are evaluated: 1) a persistent connectivity scenario to assess goodput, latency, and fairness under long delays and high loss rates, and 2) an intermittent link scenario to evaluate robustness and efficiency in a store-and-forward environment. In all experiments, node D initiated the connection and requested data from node A, resulting in data transfer along the interplanetary downlink.

The performance of our strategy, implemented in the PEPspace prototype proxying both TCP-Cubic and \textit{quic-go}-NewReno, is benchmarked against a comprehensive set of existing solutions tailored to each scenario.
For the persistent connectivity scenario, we compare against:
\begin{enumerate}
\item \textbf{End-to-End Protocols}: Standard TCP with Cubic and BBRv1 CCAs, and PicoQUIC with the latest BBRv3 algorithm, which is a C-based QUIC implementation that has demonstrated excellent performance in GEO satellite communication scenarios \cite{GEOPico}.
\item \textbf{PEP Solutions}: Three mainstream open-source PEPs are considered. PEPsal, an integrated PEP, is deployed at node B. Kcptun and PEPesc, which adopt a distributed design similar to PEPspace, are deployed at both nodes B and C. To ensure a fair comparison, Kcptun's compression is disabled, and its socket buffers and initial windows are sized to $20$ MB and $3700$ packets respectively, based on the link's BDP. 
Two configurations are evaluated for PEPesc. In PEPesc-Probed, a $30$-second idle period is introduced before data transfer to establish a near-optimal initial bandwidth estimate; in PEPesc-Immediate, data transfer starts immediately to model a more realistic case in IPNs.

\end{enumerate}
For the intermittent-link scenario, the benchmark is:
\begin{enumerate}
\item \textbf{DTN architecture:} We employ the BP/LTP protocol stack as provided by NASA's ION implementation. For comparison, PEPspace implements a store-and-forward gateway at node B, following the design paradigm in \cite{IPArchitecture}. The gateway runs as a user-space daemon that intercepts IP packets destined for an offline link, buffers them in memory, and forwards them once connectivity is restored. Functionally, this behavior is analogous to the contact plan in ION.
\end{enumerate}

\subsection{Results in the Persistent Connectivity Scenario}
\begin{figure}[t]
    \centering
    \includegraphics[width=0.485\textwidth]{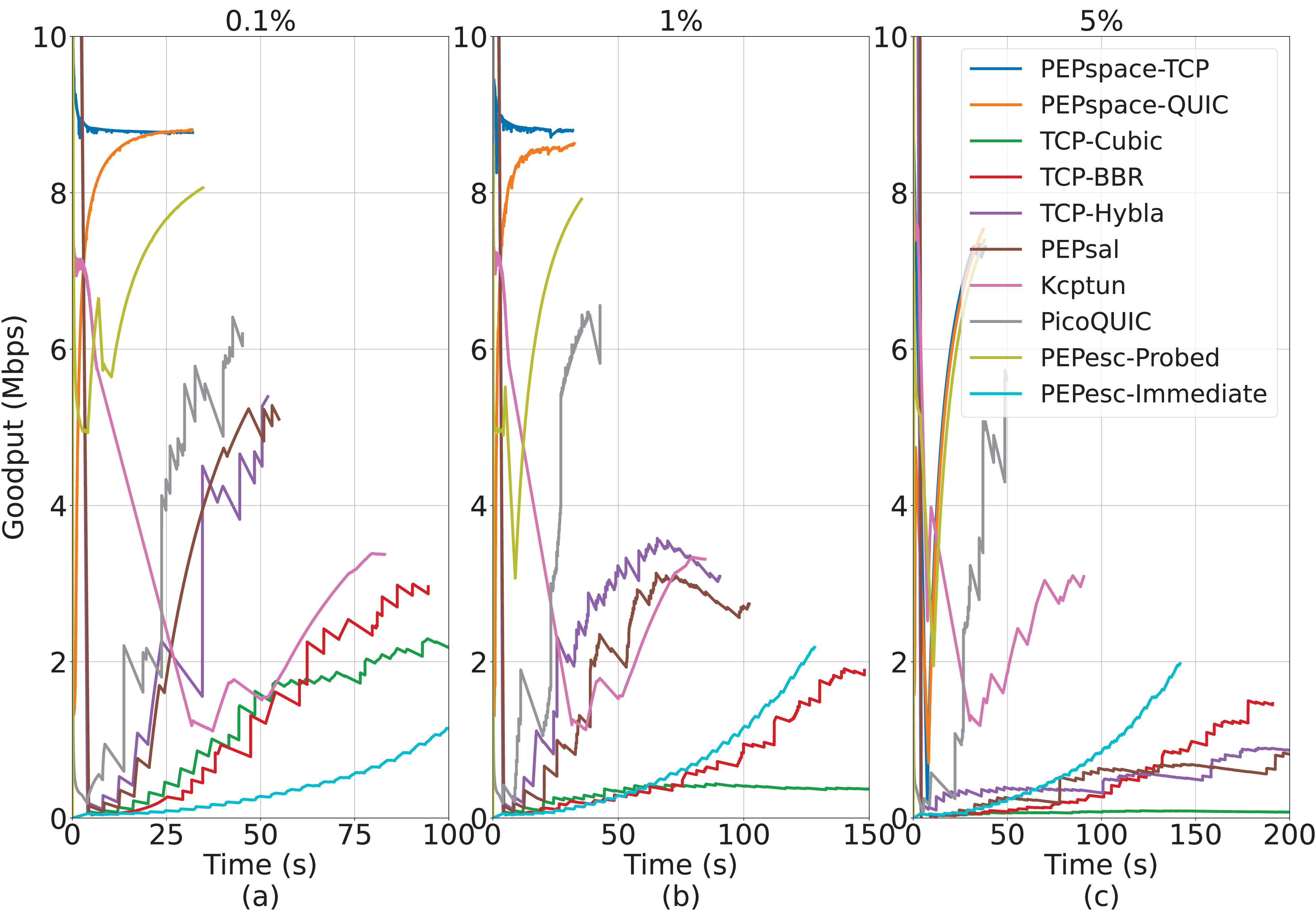}
    \caption{Goodput of 10 schemes under varying packet loss rates for a 35 MB file transfer.}
    \label{LossTp}
\end{figure}

Fig. \ref{LossTp} illustrates the goodput achieved by ten schemes during a $35$ MB file transfer under packet loss rates of $0.1\%$, $1\%$, and $5\%$. Goodput, which measures the in-order data rate received at the application layer, is given by $goodput(t_{r})=\frac{B_{r}}{t_{r}}$, where $B_{r}$ denotes the number of in-order bytes received by time $t_{r}$. The results indicate that PEPspace consistently achieved the highest goodput across all loss rates. As expected, the goodput of all schemes decreased with increasing loss rates. Specifically, when the loss rate rose from $0.1\%$ to $1\%$, the goodput of PEPspace-TCP and PEPspace-QUIC decreased by $2\%$, primarily due to the additional FEC overhead required for loss recovery. At a $5\%$ loss rate, PEPspace exhibited a temporary goodput dip at connection startup, caused by decoding delays associated with insufficient initial redundancy ratio. This behavior is consistent with the expected operation of the adaptive coding mechanism: the initial redundancy ratio is estimated from historical observations, and when the actual loss rate exceeds this estimate, the decoder experiences a short waiting period. Once ACKs conveying loss information are received, the sender promptly increases the redundancy ratio, allowing goodput to recover rapidly.

Among the comparison schemes in Fig. \ref{LossTp}, TCP-Cubic achieves the lowest goodput. Its poor performance stems from two factors: the long delay of the deep-space link, which restricts the growth of its CWND, and its overly aggressive backoff in response to packet loss. PEPsal, which adopts the same CCA as TCP-Hybla, optimizes the window growth rate and achieves goodput of $5.101$ Mbps and $5.392$ Mbps at $0.1\%$ loss, respectively. However, since Hybla still interprets loss as congestion, both PEPsal and TCP-Hybla suffer significant degradation at $1\%$ and $5\%$ loss. In contrast, Kcptun exhibits higher resilience to packet loss due to its Reed–Solomon coding. Both TCP-BBR and PicoQUIC employ the BBR algorithm, but PicoQUIC exhibits a more aggressive sending strategy, resulting in significantly faster convergence than TCP-BBR. Owing to the pre-estimated initial available bandwidth, PEPesc-Probed achieves goodput that is second only to PEPspace. By contrast, PEPesc-Immediate exhibits the same issue of prolonged bandwidth probing time as TCP-BBR, leading to a slower convergence and reduced goodput.

\begin{table}[t]
	\centering
	\caption{COMPARISON OF COV; OBSERVATION WINDOW LENGTH $1$ SECOND.}
    \renewcommand{\arraystretch}{}
	\begin{tabular}{c|c|c|c}
		\hline
		\textbf{Scheme} & \textbf{$0.1\%$} & \textbf{$1\%$} & \textbf{$5\%$}       \\ \hline\hline
		\textbf{PEPspace-TCP} &  5.567  & 5.567  & 6.164   \\ \cline{1-1}
		\textbf{PEPspace-QUIC}  & 5.567 &  5.744 & 6.082  \\ \cline{1-1}
		\textbf{TCP-BBR}  & 9.695 & 12.165 & 13.784  \\ \cline{1-1}
		\textbf{TCP-Cubic}  &  11.357 &  27.386 &  46.743 \\ \cline{1-1}
		\textbf{TCP-Hybla}  & 7.141 & 9.486 & 17.291  \\ \cline{1-1}
		\textbf{PEPsal}  &  7.348 & 10.099 & 17.776  \\ \cline{1-1}
		\textbf{Kcptun}  &  9.055 & 9.165 & 9.486  \\ \cline{1-1}
		\textbf{PicoQUIC} & 6.708   & 6.480  & 6.999 \\ \cline{1-1}
		\textbf{PEPesc-Probed} & 5.831   & 5.916  & 6.083 \\ \cline{1-1}
		\textbf{PEPesc-Immediate} & 11.402   & 11.313  & 11.874 \\ \hline\hline
	\end{tabular}
	\label{tab1}
\end{table}

To quantify goodput smoothness, the Coefficient of Variation (CoV) is adopted \cite{COV}, defined as $CoV = \frac{\sqrt{E_{t}\{goodput^{2}(i)\}-(E_{t}\{goodput(i)\})^{2}}}{E_{t}\{goodput(i)\}}$, where $goodput(i)$ represents the goodput observed within $i$-th observation window (OW), and $E_{t}\{\cdot\}$ denotes the expectation over all windows. Table~\ref{tab1} lists the CoV values corresponding to the results in Fig. \ref{LossTp}. In all scenarios, PEPspace achieves the lowest CoV, demonstrating the highest degree of goodput stability.

\begin{figure}[t]
    \centering
    \includegraphics[width=0.485\textwidth]{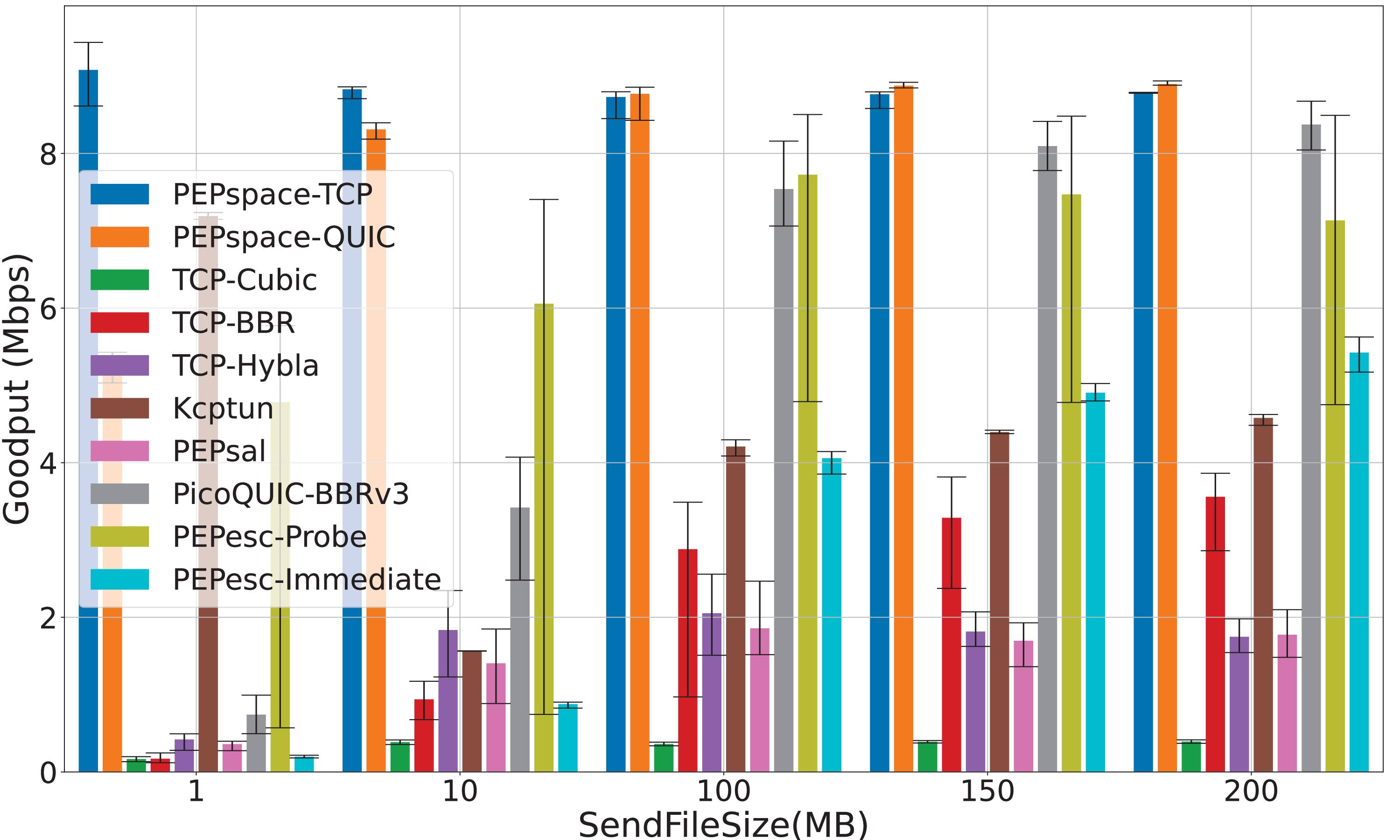}
    \caption{Goodput achieved by 10 schemes for varying file sizes at $1\%$ loss rate.}
    \label{SendFileSize}
\end{figure}

Given the long-delay characteristics of deep-space links, protocol convergence may require more time compared with terrestrial links. Fig. \ref{SendFileSize} presents the goodput achieved by the eight schemes at a $1\%$ loss rate with varying file sizes. The results show that PEPspace maintains the highest goodput across all file sizes. For a $1$ MB file transfer, PEPspace-QUIC achieves slightly lower goodput than PEPspace-TCP, since the Cubic algorithm used in TCP outperformed the NewReno algorithm employed by \textit{quic-go}. This suggests that PEPspace’s convergence performance is largely determined by the access-link protocol, decoupled from the harsh deep-space link conditions. Other protocols exhibit distinct behaviors. TCP-Hybla and PEPsal achieve higher goodput than TCP-BBR in small-file transfers, but are outperformed for files larger than $100$ MB. The goodput of both TCP-BBR and PEPesc-Immediate is positively correlated with the transfer file size, a trend indicative of a prolonged bandwidth probing phase in IPNs. Kcptun exhibits a “decline–then-rise” trend in goodput with increasing file size, a limitation inherent to its underlying KCP protocol. Its large preset send window is quickly reduced, causing goodput to stabilize around $4.5$ Mbps. 
Notably, PicoQUIC and PEPesc-Probed emerge as strong competitors. In the $200$ MB test, their maximum goodput reach $8.493$ Mbps and $8.676$ Mbps, respectively, which are second only to those of PEPspace-TCP ($8.790$ Mbps) and PEPspace-QUIC ($8.938$ Mbps). However, this impressive throughput comes at a significant cost: as discussed later, their aggressive transmission behavior leads to severe link congestion, bufferbloat, and substantial queueing delays.

\begin{figure}[ht]
    \centering
    \includegraphics[width=0.485\textwidth]{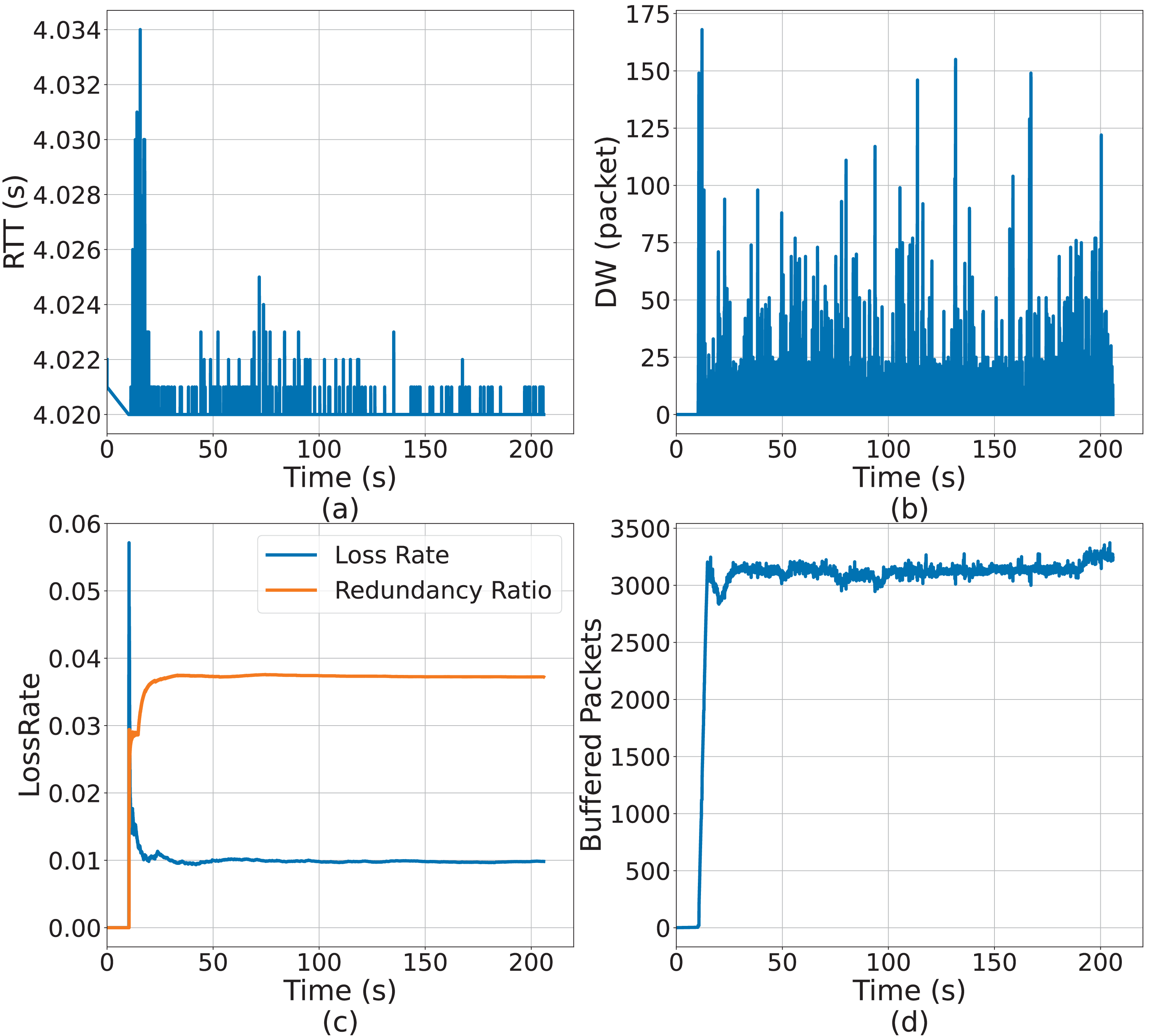}
    \caption{Overview of PEPspace performance during a $200$ MB transfer ($1\%$ loss): (a) RTT between PEPspace instances at nodes B and C; (b) DW width on node C; (c) estimated packet loss rate and redundancy ratio on node C; (d) decoder buffer on node C.}

    \label{PEPdetail}
\end{figure}

Fig. \ref{PEPdetail} further presents detailed performance indicators for PEPspace during a $200$ MB transfer at a $1\%$ loss rate. As shown in Fig. \ref{PEPdetail}(a), PEPspace maintains an RTT only slightly above the physical link RTT of $4.02$ seconds, demonstrating its ability to fully utilize bandwidth without inducing congestion. Fig. \ref{PEPdetail}(b) shows that the DW width remained consistently small, explaining the smooth goodput observed. Each lost packet could be recovered within $175$ subsequent packet arrivals, significantly mitigating HoL blocking. Fig. \ref{PEPdetail}(c) compares the estimated packet loss rate and redundancy ratio at the receiver, showing that the adaptive coding mechanism effectively selects appropriate redundancy ratios to balance low-latency recovery with bandwidth overhead. Fig. \ref{PEPdetail}(d) depicts the number of packets buffered at the SC decoder. The bottleneck link’s BDP is calculated as $\frac{10 \text{ Mbps} \times 4.02 \text{ s}}{1500 \times 8 \text{ bits/byte}} \approx 3350 \text{ packets}.$ This implies that the SC decoder needs to buffer approximately one BDP worth of packets to decode effectively. Such buffering overhead is one of primary costs of the SC mechanism in exchange for zero retransmissions and low-latency loss recovery, representing a tradeoff between memory consumption and performance.

\begin{figure}[t]
    \centering
    \includegraphics[width=0.485\textwidth]{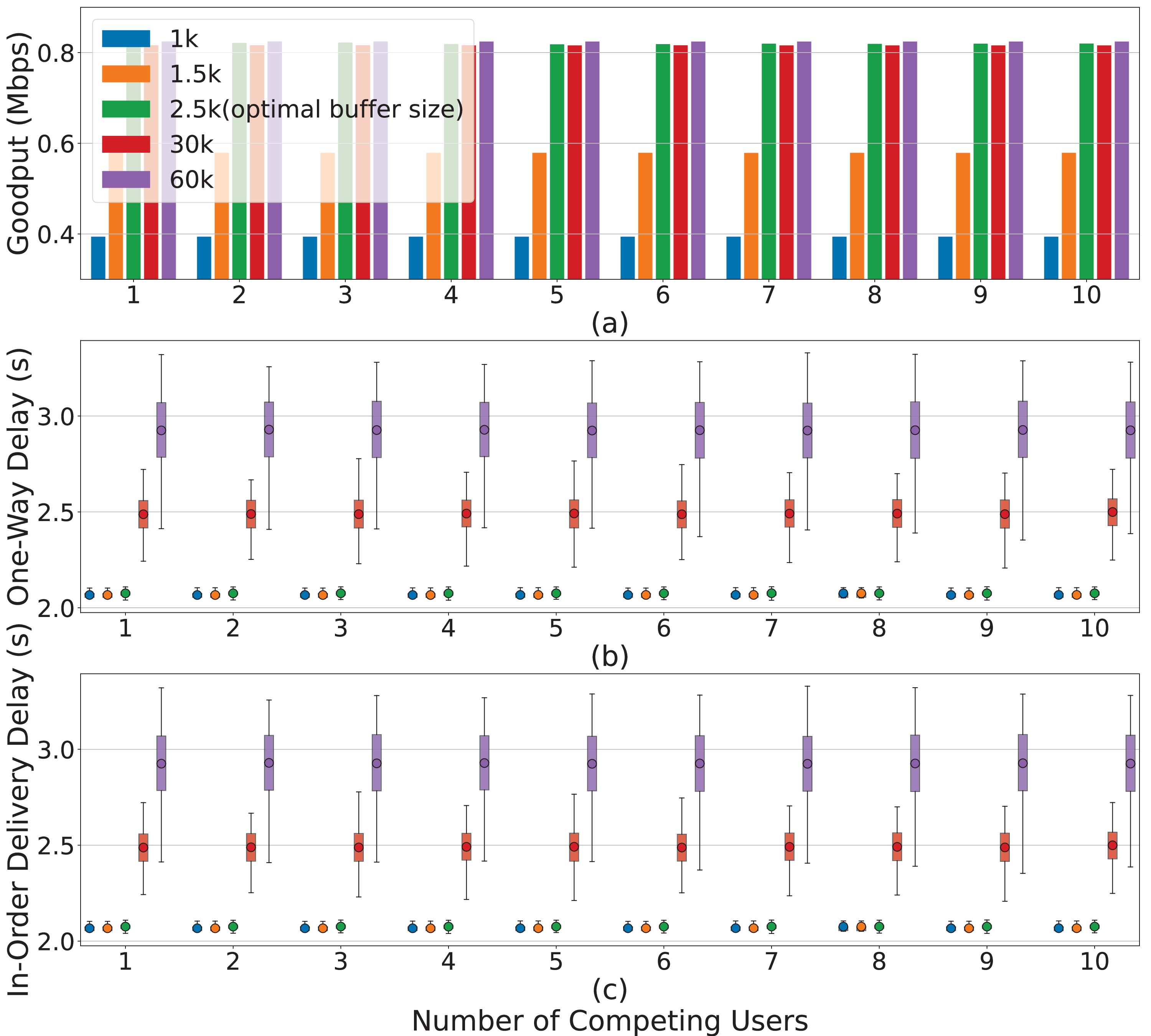}
    \caption{(a) Per-flow goodput, (b) one-way delay and (c) in-order delivery delay for buffer sizes of 1 KB, 1.5KB, 2.5 KB, 30 KB and 60 KB in a 10-flow scenario.}
    \label{Buffer}
\end{figure}

Next, the performance in a multi-flow coexistence scenario is evaluated. In this scenario, multiple competing flows are established between nodes A and D, with each flow transferring up to $35$ MB of data and a bottleneck link loss rate of $1\%$. First, the correctness of the optimal buffer size derived in Section \ref{KOPT} is validated in a scenario with $10$ competing flows. According to Eq. \ref{EQKPOT}, the optimal flow-control buffer per flow is $K_{\mathrm{opt}} = \frac{10\ \mathrm{Mbps}\times 20\ \mathrm{ms}}{10\times 8} = 2.5\ \mathrm{KB}$. Fig. \ref{Buffer} shows the per-user goodput, one-way delay, and in-order delivery delay when PEPspace proxied quic-go, with each user being allocated a flow control buffer of $1$ KB, $1.5$ KB, $2.5$ KB, $30$ KB, and $60$ KB, respectively. For a packet $p$, let $t_{\text{send}}(p)$ denote its transmission time and $t_{\text{arrive}}(p)$ its arrival time at the receiver. The OWD is defined as $OWD(p) = t_{arrive}(p) - t_{send}(p)$, which measures the total delay experienced by a packet in the network. The in-order delivery delay is the time elapsed from a packet's first send until its in-sequence delivery to the application, which reflects the end-to-end delay perceived by the application.

As shown in Fig. \ref{Buffer}(a), when the buffer size is smaller than $2.5$ KB, goodput increases with buffer size. This behavior corresponds to the under-buffered regime, where the flow-control mechanism constrains the sender rate and the downstream $10$ Mbps bottleneck is not fully utilized. When the buffer exceeds $2.5$ KB, no further goodput improvement is observed. Instead, as seen in Fig. \ref{Buffer}(b), OWD increases approximately linearly with buffer size. This corresponds to the over-buffered regime in which any additional buffer beyond $K_{opt}$ only accumulates packets at the proxy and increases queuing delay without improving goodput. Fig. \ref{Buffer}(c) shows that the in-order delivery delay for each user is nearly identical to the OWD. This indicates that the recovery mechanism based on SC can repair packet losses with extremely low HoL blocking delay, allowing the application-perceived delivery delay to closely track the true network latency. By contrast, retransmission-based protocols typically incur at least $1.5$ RTTs of additional in-order delivery delay when losses occur.

\begin{figure*}[t]
    \centering
    \includegraphics[width=0.98\textwidth]{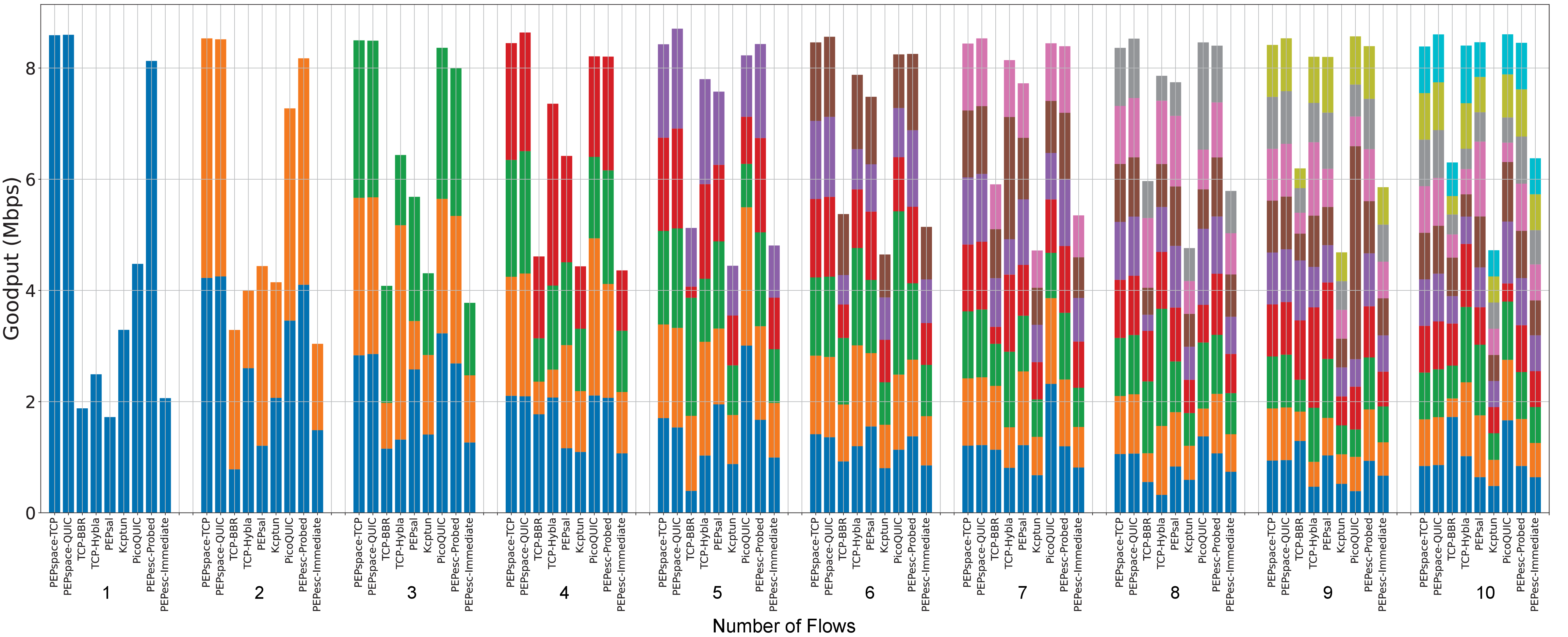}
    \caption{Goodput comparison for multiple flows with packet loss rate $1\%$.}
    \label{tp35}
\end{figure*}

Fig. \ref{tp35} uses stacked bar charts to display the goodput performance of the eight schemes in scenarios with $1$ to $10$ flows. The total height of each bar represents the aggregate goodput of all flows, while the height of each colored sub-bar corresponds to the goodput of a single flow at the moment the first flow completed its $35$ MB transfer. The results show that PEPspace achieves near-capacity aggregate goodput in all multi-flow scenarios, with an average of $8.455$ Mbps and $8.571$ Mbps when proxying TCP and QUIC, respectively. In contrast, the aggregate goodput of TCP-Cubic, TCP-BBR, TCP-Hybla, PEPsal and PEPesc-Immediate increases with the number of flows. TCP-Hybla and PEPsal achieve maximum aggregate goodputs of $8.403$ Mbps and $8.463$ Mbps, respectively. The link utilization of Kcptun, however, remained at approximately $50\%$ in all scenarios, and did not improve as the number of flows increased. This is because Kcptun maps multiple competing flows to a single underlying Kcp stream, preventing an increase in flow count from improving link utilization. PEPesc-Probed exhibits aggregate goodput comparable to that of PEPspace, and achieves a bandwidth utilization of over $80\%$ across all flows. Furthermore, when the number of flows exceeded three, the aggregate goodput of PicoQUIC approached that of PEPspace. This improvement is attributed to the reduced bandwidth probing time in multi-flow settings, which mitigates PicoQUIC’s single-flow probing delay.

\begin{figure}[h]
    \centering
    \includegraphics[width=0.485\textwidth]{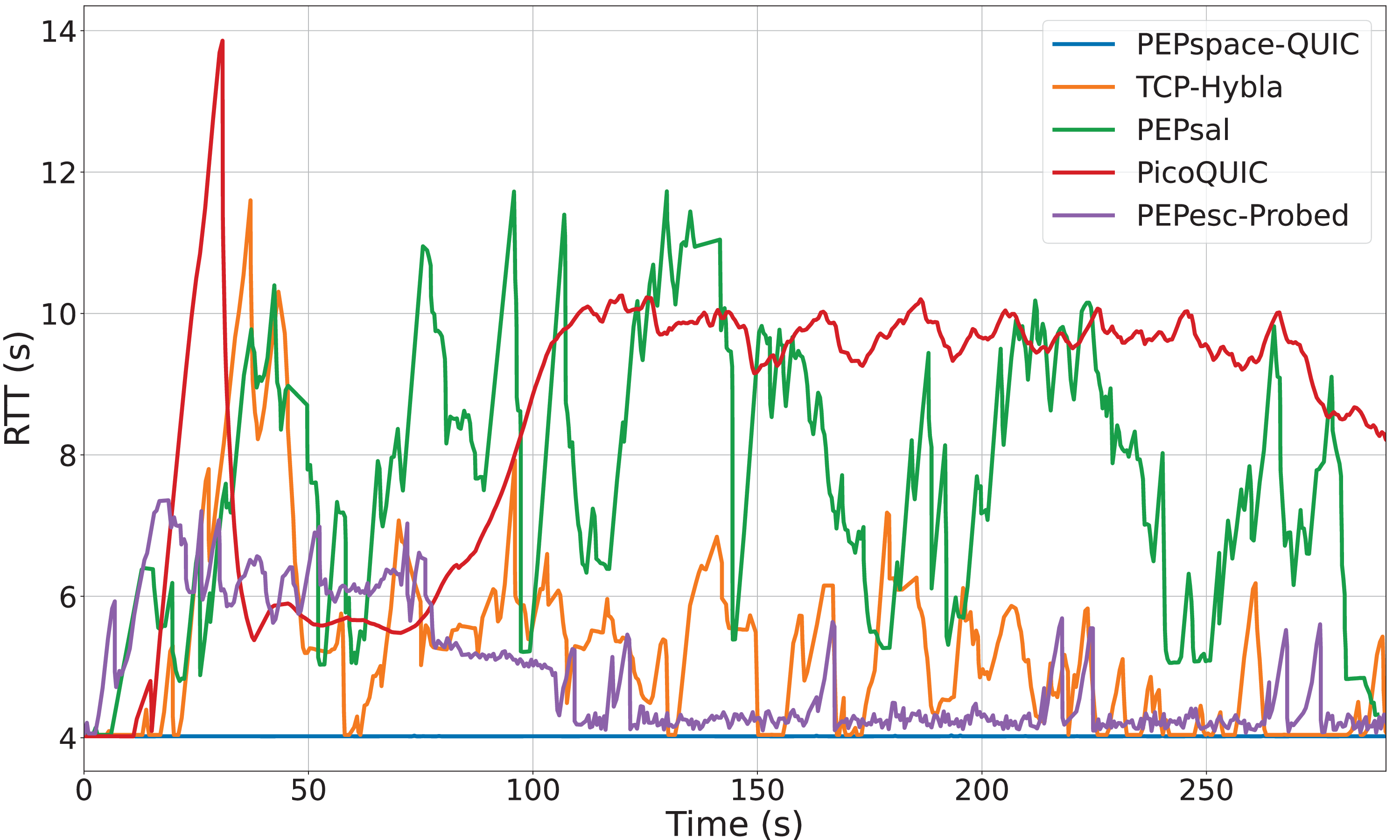}
    \caption{RTT measured by UDP ping in a $10$-flow scenario.}
    \label{RTT}
\end{figure}

It should be noted that the aggressive sending strategies of TCP-Hybla, PEPsal, PicoQUIC and PEPesc-Probed in multi-flow experiments may lead to severe link congestion, as shown in Fig. \ref{RTT}. To quantify link congestion, an independent UDP Ping program is initiated at the start of the transfer, sending a $13$-byte Ping packet every $0.5$ seconds to calculate the link RTT. Fig. \ref{RTT} presents the measured RTT for PEPspace-QUIC, TCP-Hybla, PEPsal, PicoQUIC and PEPesc-Probed in the $10$-flow scenario. The results reveal that the comparison schemes produced sustained high RTTs indicative of continuous congestion. This indicates that while they can achieve near-capacity aggregate goodput in some scenarios, it comes at the cost of increased network latency and severe congestion. In contrast, PEPspace-QUIC maintained a stable RTT close to the physical delay while sustaining high goodput, enabling efficient data transfer without degrading network stability.

\begin{figure}[h]
    \centering
    \includegraphics[width=0.485\textwidth]{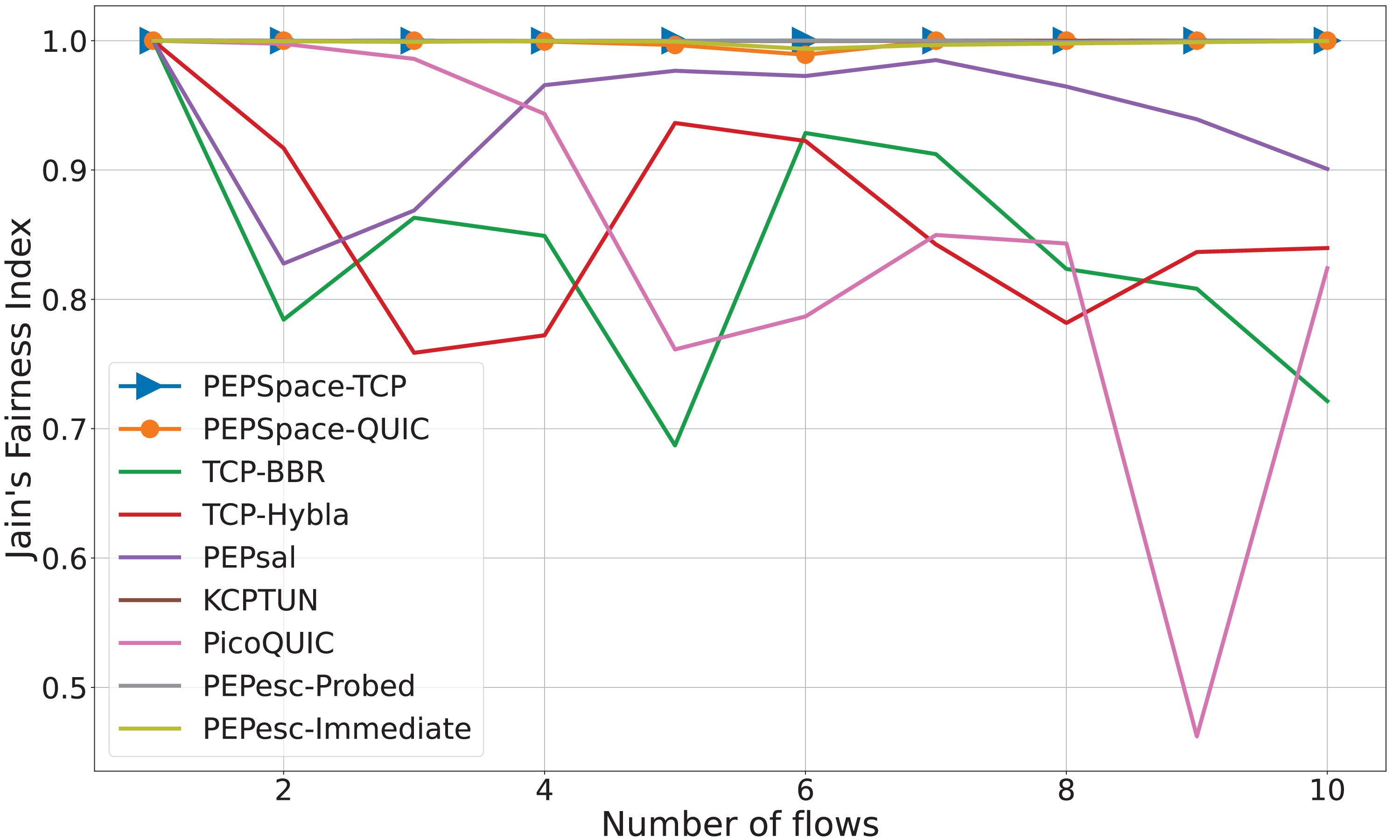}
    \caption{Jain's fairness index for eight schemes in a $10$-flow scenario.}
    \label{Jians}
\end{figure}

\begin{figure}[h]
    \centering
    \includegraphics[width=0.485\textwidth]{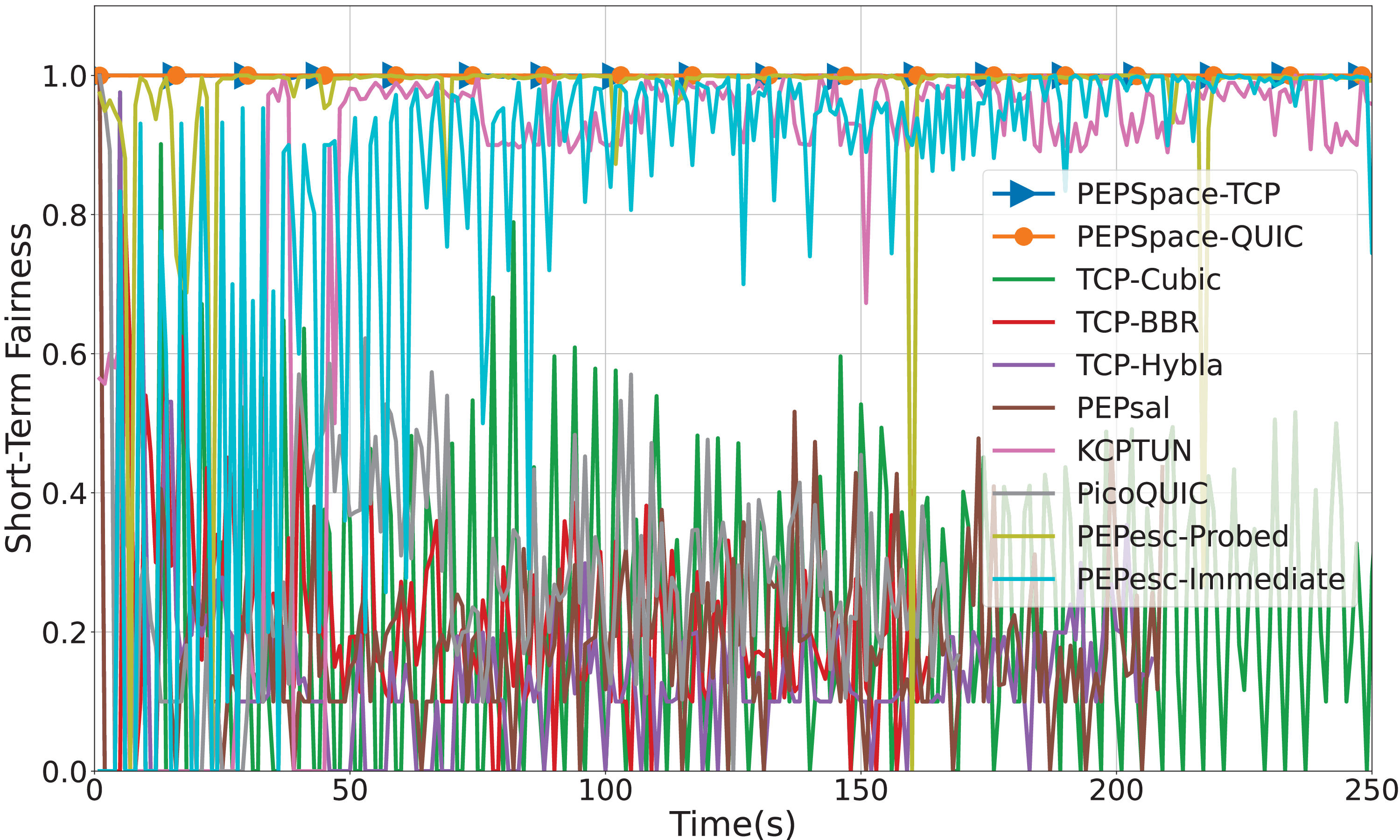}
    \caption{SFI over time for eight schemes in a $10$-flow scenario.}
    \label{ShortIain}
\end{figure}

Jain’s fairness index for the multi-flow scenario illustrated in Fig. \ref{tp35} is compared across schemes in Fig. \ref{Jians}. Both PEPspace-TCP and PEPspace-QUIC exhibit  excellent fairness, with fairness indices exceeding $0.98$ in all tested scenarios. This indicates that PEPspace, as an effective proxy solution, satisfies the requirement of ensuring that intercepted connections share the bottleneck link bandwidth equitably. To further evaluate per-flow fairness over time, the Short-term Fairness Index (SFI) \cite{SFI} is introduced. SFI evaluates each individual flow’s goodput fairness during the transmission process and is defined as the Jain’s fairness index within OWs: $\mathrm{SFI} =  E_{t}\left\{\frac{\big(\sum_{l=1}^{L}\mathrm{goodput}_{l}(t)\big)^{2}}{L\cdot\sum_{l=1}^{L}\mathrm{goodput}_{l}^{2}(t)}\right\}$ for $L$ flows,
where $\mathrm{goodput}_{l}(t)$ denotes the goodput of flow $l$ in OW $t$. Fig. \ref{ShortIain} compares the SFI traces computed over $1$-second OWs for the $10$-flow case. PEPspace’s SFI remained approximately $1$ throughout the transfer, indicating near-ideal temporal fairness. Taken together, Figs. \ref{Jians} and \ref{ShortIain} show that PEPspace not only attains fair aggregate goodput allocation but also preserves per-flow fairness across time during the transfer.

\subsection{Robustness to Transient Interruptions}
\begin{figure}[t]
    \centering
    \includegraphics[width=0.485\textwidth]{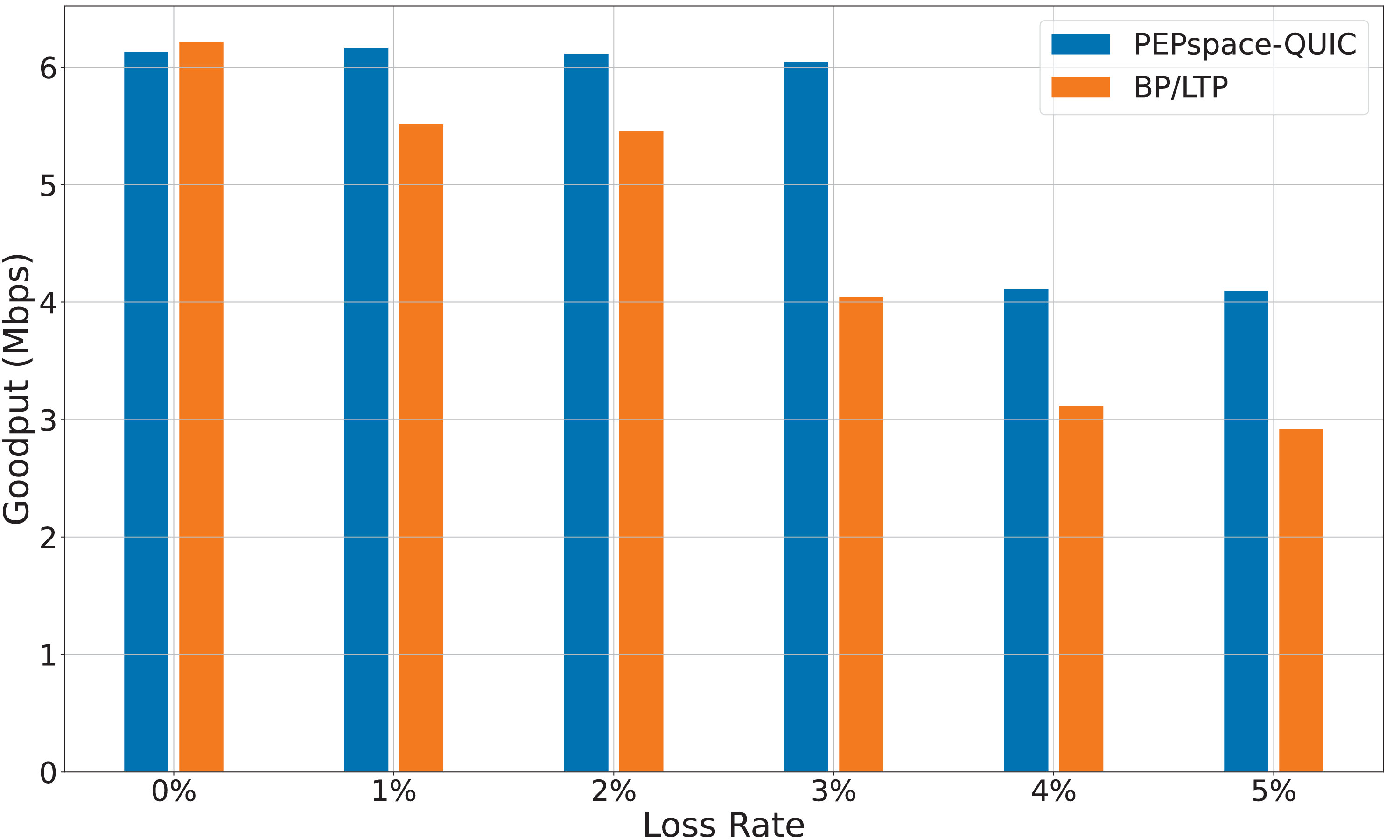}
    \caption{Goodput of PEPspace and BP/LTP for a $35$-MB file transfer under a $12$-s disruption at different loss rates.}
    \label{InterruptTP}
\end{figure}

We position this comparison within our target Earth-Moon scenario, which is characterized primarily by transient interruptions (e.g., seconds-long handovers or brief signal fades) rather than the long-duration (hours to days) disruptions addressed by deep-space DTN architectures. Specifically, Fig. \ref{InterruptTP} presents the performance of 35 MB file transfers under a 12-second network interruption with different packet loss rates. For BP/LTP, the ION Contact Plan is employed to define link availability windows, and the \textit{bpdriver} tool is used to generate $35$ bundles of $1$ MB each. The results indicate that PEPspace experienced a noticeable goodput reduction compared with the uninterrupted scenario. In the loss-free case, BP/LTP achieved slightly higher goodput than PEPspace, primarily due to the additional FEC overhead incurred by PEPspace. However, in lossy scenarios, the performance of the two schemes diverged significantly. As shown in Fig. \ref{InterruptTP}, BP/LTP goodput decreased steadily as the loss rate increased, as its underlying LTP protocol depends on ARQ-based recovery. In long-delay deep-space links, this ARQ mechanism severely reduces link utilization. By contrast, PEPspace goodput remained largely insensitive to packet loss, with a significant decline observed only beyond a $3\%$ loss rate. This behavior can be attributed to insufficient pre-configured redundancy ratio, which necessitated adaptive adjustment upon receiving feedback, consistent with the observations reported in Fig. \ref{LossTp} for the uninterrupted scenario.

\begin{figure}[t]
    \centering
    \includegraphics[width=0.485\textwidth]{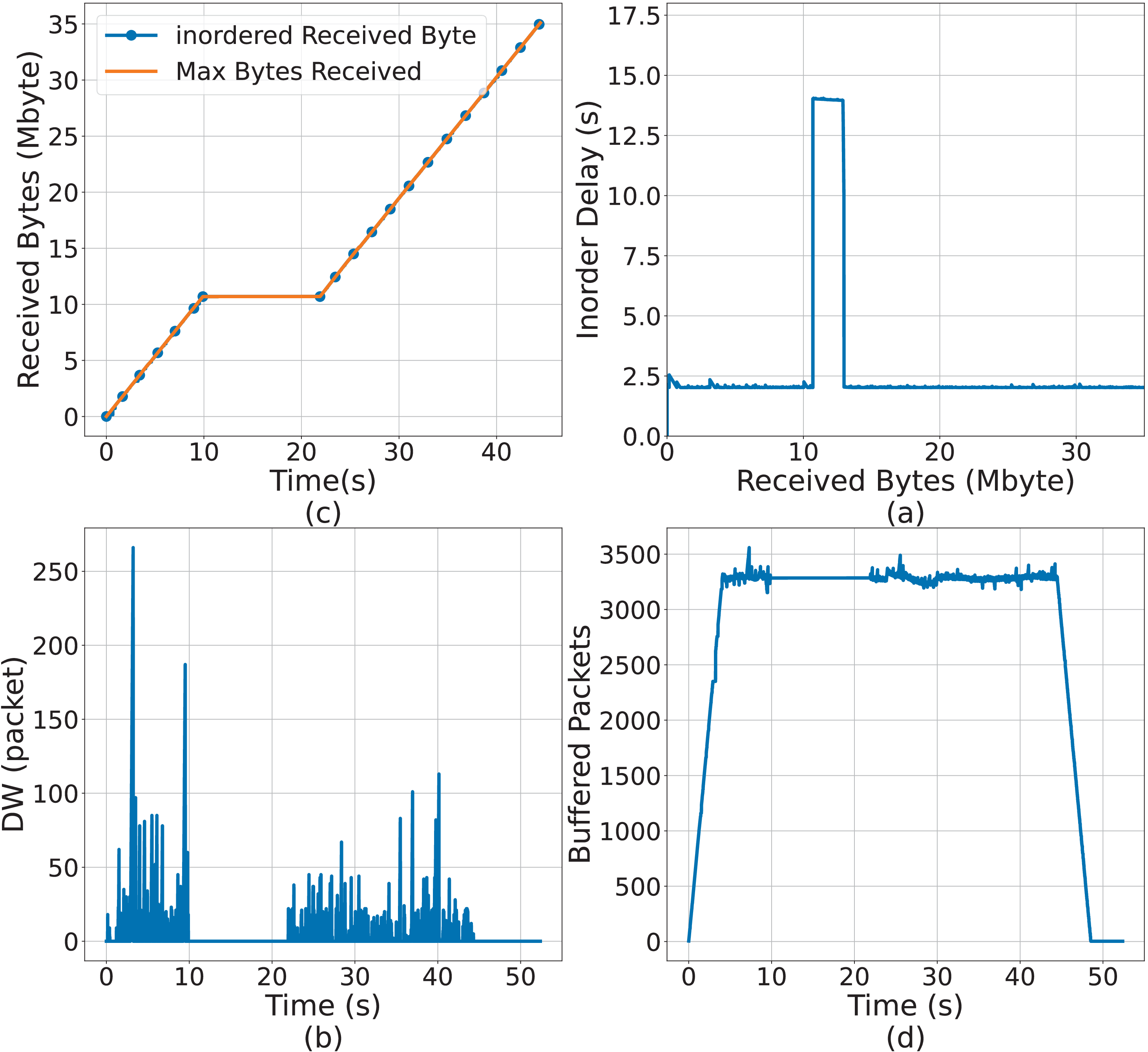}
    \caption{Overview of PEPspace-QUIC under a $12$-s disruption with $1\%$ packet loss: (a) cumulative in-order and total received bytes, (b) in-order delivery delay, (c) DW width, and (d) decoder buffer.}
    \label{Interruptdetail}
\end{figure}

To further explain the impact of interruptions, Fig. \ref{Interruptdetail} presents detailed performance indicators for PEPspace-QUIC at a $1\%$ loss rate. Fig. \ref{Interruptdetail}(a) plots the cumulative in-order bytes received alongside the total received bytes. The curve exhibits a plateau of approximately $12$ seconds, clearly illustrating how the interruption prolonged the overall transfer time and thereby reduced average goodput. Fig. \ref{Interruptdetail}(b) shows the effect of disruption on end-to-end in-order delivery delay: packets transmitted during the disruption exhibited delays increased by roughly $12$ seconds. Figs. \ref{Interruptdetail}(c) and \ref{Interruptdetail}(d) demonstrate that, under the IP-based store-and-forward strategy described in \cite{IPArchitecture}, the disruption can be interpreted as a transmission pause rather than a loss event. Consequently, its impact on DW and decoder buffer is negligible.

\section{Discussion: Towards a Unified IP/DTN Transport Architecture}
The evolution of IPN has sparked a significant debate within the research community, revolving around two competing paradigms: (i) deep-space IP architectures advocated by groups such as TIPTOP, and (ii) DTN architectures centered on the BP/LTP protocol suite. The deep-space IP approach offers the advantage of seamless integration with the vast terrestrial Internet ecosystem, allowing for the direct reuse of applications and protocols. However, as demonstrated in this paper, the standard IP stack requires significant enhancements to cope with the extreme conditions of deep-space links. In contrast, DTN architectures exhibit superior robustness against long and unpredictable link disruptions, but this robustness sacrifices interoperability: cross-architecture communication demands complex protocol translation, which hinders seamless end-to-end communication.

We argue that the future of the IPN is not a choice between IP and DTN, but their convergence. In environments where latency and disruptions are relatively manageable, such as the Earth-Moon network, deep-space-optimized IP stacks can directly support diverse network demands. In harsher environments, such as Mars missions, the DTN architecture remains indispensable for ensuring reliability. The deeper value of the proposed NTSP architecture lies precisely in this bridging capability: NTSP is not merely a secure acceleration mechanism for IP-based transport but also a foundational framework for a unified IP/DTN transport architecture. By virtue of its secure connection-splitting and modular transport-layer design, NTSP can safely terminate IP connections and redirect payloads to alternative transport mechanisms such as BP/LTP through architectural abstraction.

To ground our discussion and explore the practical challenges of such a hybrid architecture, we developed a proof-of-concept (PoC) implementation within the NTSP framework. In this prototype, the IPN-aware QUIC tunnel is replaced by an aggregation layer interfacing with the BP/LTP stack. Standard IP traffic is thus redirected to the DTN stack, forming a complete data path of “TCP/QUIC → BP/LTP → TCP/QUIC.” To isolate core technical issues, the PoC adopts a minimalistic design consisting of the following key functional components:
\begin{enumerate}
\item \textbf{Endpoint Termination}: Following the NTSP architecture, proxy entities terminate incoming TCP/QUIC connections from the source client.
\item \textbf{Aggregation Buffer}: Incoming stream data are collected and sorted; once a predefined threshold ($1$ MB) or timeout ($800$ ms) is reached, data blocks are encapsulated into a single bundle for transmission.
\item \textbf{Reordering Buffer}: Since BP does not guarantee in-order delivery, received bundles are temporarily stored and reordered according to sequence numbers before being written into the outbound TCP/QUIC stream toward the destination server.
\item \textbf{DTN Stack Interface}: The PoC interfaces with the ION implementation to inject bundles into the local BP agent for transmission and to receive them at the destination.
\end{enumerate}

\begin{figure}[h]
    \centering
    \includegraphics[width=0.485\textwidth]{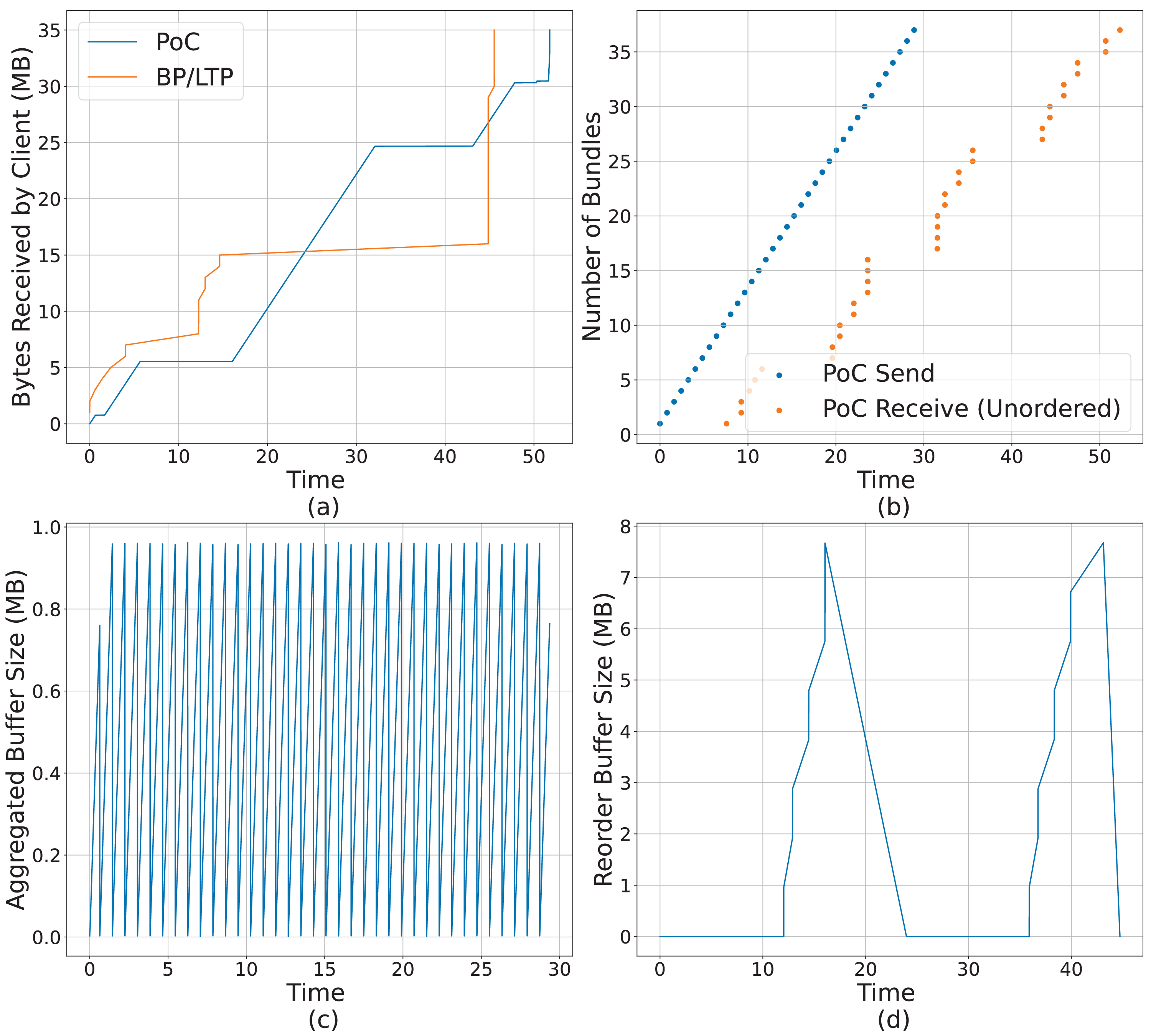}
    \caption{Overview of PoC for a $35$-MB file transfer under a $1\%$ loss rates.}
    \label{PEPDTN}
\end{figure}

Fig. \ref{PEPDTN} presents the PoC evaluation results for a single TCP flow in the Earth-Moon network scenario with a $1\%$ loss rate, as detailed in Section IV. Fig. \ref{PEPDTN} (a) plots the cumulative data received by the client. The successful and complete transfer of the entire file confirms the basic feasibility of integrating the BP/LTP stack within the NTSP framework. The horizontal gap between the curves is not due to implementation inefficiency but arises from the inherent semantic mismatch between byte-stream protocols and message-oriented protocols. Fig. \ref{PEPDTN}(b) depicts the timing distribution of bundle transmission and reception, while Fig. \ref{PEPDTN}(c) and Fig. \ref{PEPDTN}(d) show the variations in the aggregation and reordering buffer sizes, respectively. These results confirm the basic feasibility of integrating the BP/LTP stack within the NTSP framework.

While conceptually feasible, our PoC further reveals several fundamental research challenges at the interface between the IP and DTN paradigms:
\begin{enumerate}
\item \textbf{Performance Trade-offs Induced by Semantic Mismatch}: The intrinsic differences between TCP/QUIC’s byte-stream semantics and BP’s message semantics lead to unavoidable performance trade-offs. To generate reasonably sized BP bundles and avoid excessive overhead from small bundles, the aggregation buffer must accumulate sufficient data, introducing serialization delay detrimental to latency-sensitive or interactive applications. Furthermore, the reordering buffer must wait for out-of-order bundles before reconstructing the original ordered byte stream; a single delayed or lost bundle may cause HoL blocking, stalling subsequent delivery. This delay stems from the fundamental mismatch between stream-oriented and message-oriented abstractions. A key research question thus arises: How can segmentation, scheduling, and reassembly schemes be redesigned to minimize delay overhead while retaining DTN’s reliability guarantees?
\item \textbf{QoS Scheduling under Concurrent Flows}: Our PoC is limited to a single flow, whereas real-world systems must support multiple concurrent flows with diverse QoS requirements. A core scheduling challenge emerges: How should multiple flows be aggregated into discrete bundles according to their QoS priorities? A simple First-In-First-Out (FIFO) approach can easily cause QoS inversion and fairness issues. A critical research direction, therefore, is to develop QoS-aware scheduling and aggregation strategies that balance inter-flow fairness with priority differentiation when multiplexing multiple streams into unified BP flows.
\item \textbf{Application-Layer Gatewaying for Native Endpoints}: Beyond tunneling, a truly unified architecture must support communication between IP-native applications and DTN-native endpoints. This requires the NTSP entity to function as an application-layer gateway (e.g., mapping an HTTP request to a corresponding bundle payload and metadata). Designing generic and efficient mechanisms for this stateful, application-specific translation, while preserving security, remains a major open research area.
\end{enumerate}

These challenges motivate clear avenues for future research. Our long-term vision is an intelligent, policy-driven NTSP architecture that adaptively selects forwarding strategies for heterogeneous network environments. By demonstrating the feasibility of a hybrid architecture and exposing remaining open problems, this work both demonstrates a high-performance approach to IP-based deep-space networking and lays the groundwork for a more unified, resilient, and adaptive Interplanetary Internet.

\section{Conclusion}
This paper presents a secure transport acceleration strategy to address the performance degradation of TCP and QUIC in IPNs characterized by extreme delay, high loss, and frequent disruptions. Our approach is founded on the NTSP architecture, which enables secure connection splitting for end-to-end encrypted flows while preserving application-layer security. Complementing this architecture, we design an IPN-aware transport policy that integrates a rate-based congestion control algorithm with an adaptive FEC scheme for low-latency recovery. Furthermore, we introduce a theoretically grounded backpressure mechanism with an analytically derived buffer-sizing rule to prevent bufferbloat and ensure system stability.

We implement this strategy in our prototype, PEPspace, and conduct extensive evaluations in representative Earth–Moon scenarios. The results demonstrate that PEPspace consistently achieves stable, near-capacity goodput, significantly outperforming conventional TCP/QUIC variants and open-source PEPs. Critically, it avoids the severe link congestion and bufferbloat in other aggressive schemes. Moreover, PEPspace exhibits excellent fairness in multi-flow scenarios and robust performance during link interruptions, and demonstrates robust performance against transient link interruptions by leveraging IP store-and-forward capability.

Furthermore, this paper extends the discussion beyond IP-native acceleration by exploring the role of the NTSP as a foundation for a unified IP/DTN architecture. Our PoC and subsequent analysis identified a concrete set of open research questions, providing a clear roadmap for future work in creating truly adaptive and interoperable interplanetary networks.

Overall, the proposed strategy provides a practical and effective solution for accelerating encrypted transport over IPNs. Future work will incorporate more realistic deep-space channel models and extend the transport policy with differentiated QoS scheduling to support diverse mission requirements.


\bibliographystyle{IEEEtran}
\bibliography{Mybib}

@article{de2023roadmap,
  title={Roadmap for solar system research 2022},
  author={De Moortel, I and others},
  journal={White Paper, the Science and Technology Facilities Council (STFC)},
  year={2023}
}

@inproceedings{Artemis,
  title={Artemis: an overview of NASA's activities to return humans to the Moon},
  author={Creech, Steve and Guidi, John and Elburn, Darcy},
  booktitle={IEEE Aerosp. Conf. Proc.},
  pages={1--7},
  year={2022},
  organization={IEEE}
}

@inproceedings{Moonlight,
  title={The Lunar Pathfinder PNT experiment and Moonlight navigation service: The future of lunar position, navigation and timing},
  author={Giordano, Pietro and Malman, Floor and Swinden, Richard and others},
  booktitle={Proc. 2022 ION Int. Tech. Meet.},
  pages={632--642},
  year={2022}
}

@article{ChangE5,
  title={Scientific objectives and payloads of the lunar sample return mission—Chang’E-5},
  author={Zhou, Changyi and Jia, Yingzhuo and Liu, Jianzhong and others},
  journal={Adv. Space Res},
  volume={69},
  number={1},
  pages={823--836},
  year={2022},
  publisher={Elsevier}
}

@article{ChinaMars,
  title={China’s Mars exploration mission and science investigation},
  author={Li, Chunlai and Zhang, Rongqiao and Yu, Dengyun and others},
  journal={Space Sci. Rev},
  volume={217},
  number={4},
  pages={57},
  year={2021},
  publisher={Springer}
}

@article{IPNs,
  title={The sky is NOT the limit anymore: Future architecture of the interplanetary Internet},
  author={Alhilal, Ahmad and Braud, Tristan and Hui, Pan},
  journal={IEEE Aerosp. Electron. Syst. Mag.},
  volume={34},
  number={8},
  pages={22--32},
  year={2019},
  publisher={IEEE}
}

@techreport{ioagmoon,
  author      = {{Lunar Communications Architecture Working Group, Interagency Operations Advisory Group}},
  title       = {{The Future Lunar Communications Architecture, Report of the Interagency Operations Advisory Group}},
  institution = {Interagency Operations Advisory Group},
  month       = jan,
  year        = {2022},
  note        = {Available: \url{https://www.ioag.org/Public%20Documents/Lunar%20communications%20architecture%20study%20report%20FINAL%20v1.3.pdf}}
}

@techreport{ioagmars,
  author      = {{Mars and Beyond Communications Architecture Working Group, Interagency Operations Advisory Group}},
  title       = {{The Future Mars Communications Architecture, Report of the Interagency Operations Advisory Group}},
  institution = {Interagency Operations Advisory Group},
  month       = feb,
  year        = {2022},
  note        = {Available: \url{https://www.ioag.org/Public%20Documents/MBC%20architecture%20report%20final%20version%20PDF.pdf}}
}

@inproceedings{Lunanet,
  title={Lunanet: a flexible and extensible lunar exploration communications and navigation infrastructure},
  author={Israel, David J and Mauldin, Kendall D and Roberts, Christopher J and others},
  booktitle={Proc. 2020 IEEE Aerosp. Conf.},
  pages={1--14},
  year={2020},
  organization={IEEE}
}

@article{Joint,
  title={Joint Age and Coverage-Optimal Satellite Constellation Relaying in Cislunar Communications with Hybrid Orbits},
  author={Yuan, Afang and Hu, Zhouyong and Sun, Zhili and Zhang, Qinyu and Yang, Zhihua},
  journal={IEEE Trans. Commun.},
  year={2025},
  publisher={IEEE}
}

@inproceedings{lunarexpeditions,
  title={The artemis program: An overview of nasa's activities to return humans to the moon},
  author={Smith, Marshall and Craig, Douglas and Herrmann, Nicole and Mahoney, Erin and Krezel, Jonathan and McIntyre, Nate and Goodliff, Kandyce},
  booktitle={Proc. 2020 IEEE Aerosp. Conf.},
  pages={1--10},
  year={2020},
  organization={IEEE}
}

@article{lunarbases,
  title={Space microgrids for future manned lunar bases: A review},
  author={Saha, Diptish and Bazmohammadi, Najmeh and Raya-Armenta, Jos{\'e} Maurilio and others},
  journal={IEEE Open Access J. Power Energy},
  volume={8},
  pages={570--583},
  year={2021},
  publisher={IEEE}
}

@article{gibney2022asteroids,
  title={Asteroids, hubble rival and moon base: China sets space plan},
  author={Gibney, Elizabeth},
  journal={Nature},
  volume={603},
  pages={19--20},
  year={2022}
}

@article{Futurespacenetworks,
  title={Future space networks: Toward the next giant leap for humankind},
  author={Abdelsadek, Mohammed Y and Chaudhry, Aizaz U and Darwish, Tasneem and others},
  journal={IEEE Trans. Commun},
  volume={71},
  number={2},
  pages={949--1007},
  year={2022},
  publisher={IEEE}
}

@article{TCPinDSCN,
  title={Performance of TCP protocols in deep space communication networks},
  author={Akan, Ozgur B and Fang, H and Akyildiz, Ian F},
  journal={IEEE Commun. Lett.},
  volume={6},
  number={11},
  pages={478--480},
  year={2002},
  publisher={IEEE}
}

@article{linkinterruptions,
  title={Effects of solar scintillation on deep space communications: Challenges and prediction techniques},
  author={Xu, Guanjun and Song, Zhaohui},
  journal={IEEE Wirel. Commun.},
  volume={26},
  number={2},
  pages={10--16},
  year={2019},
  publisher={IEEE}
}

@article{DTN,
  title={Delay-tolerant networking: an approach to interplanetary internet},
  author={Burleigh, Scott and Hooke, Adrian and Torgerson, Leigh and others},
  journal={IEEE Commun. Mag.},
  volume={41},
  number={6},
  pages={128--136},
  year={2003},
  publisher={IEEE}
}

@misc{BPv7,
    series =    {Request for Comments},
    number =    9171,
    howpublished =  {RFC 9171},
    publisher = {RFC Editor},
    doi =       {10.17487/RFC9171},
    url =       {https://www.rfc-editor.org/info/rfc9171},
    author =    {Scott Burleigh and Kevin Fall and Edward J. Birrane},
    title =     {{Bundle Protocol Version 7}},
    pagetotal = 53,
    year =      2022,
    month =     jan,
}

@misc{LTP,
    series =    {Request for Comments},
    number =    5326,
    howpublished =  {RFC 5326},
    publisher = {RFC Editor},
    doi =       {10.17487/RFC5326},
    url =       {https://www.rfc-editor.org/info/rfc5326},
    author =    {Scott C. Burleigh and Stephen Farrell and Manikantan Ramadas},
    title =     {{Licklider Transmission Protocol - Specification}},
    pagetotal = 54,
    year =      2008,
    month =     sep,
}

@techreport{IPArchitecture,
    number =    {draft-many-tiptop-ip-architecture-01},
    type =      {Internet-Draft},
    institution =   {Internet Engineering Task Force},
    publisher = {Internet Engineering Task Force},
    note =      {Work in Progress},
    url =       {https://datatracker.ietf.org/doc/draft-many-tiptop-ip-architecture/01/},
    author =    {Marc Blanchet and Wesley Eddy and Tony Li},
    title =     {{An Architecture for IP in Deep Space}},
    pagetotal = 21,
    year =      2025,
    month =     jul,
    day =       7,
}

@INPROCEEDINGS{SRv6,
  author={Zhu, Yihan and Liu, Jun and Zhao, Kanglian and Li, Wenfeng},
  booktitle={Proc. 2024 Int. Conf. Future Commun. Netw. (FCN)}, 
  title={SRv6 Based Store-Carry-and-Forward Networking for Deep Space}, 
  year={2024},
  volume={},
  number={},
  pages={1-6},
  doi={10.1109/FCN64323.2024.10985507}}

@misc{TCP,
    series =    {Request for Comments},
    number =    9293,
    howpublished =  {RFC 9293},
    publisher = {RFC Editor},
    doi =       {10.17487/RFC9293},
    url =       {https://www.rfc-editor.org/info/rfc9293},
    author =    {Wesley Eddy},
    title =     {{Transmission Control Protocol (TCP)}},
    pagetotal = 98,
    year =      2022,
    month =     aug,
}

@techreport{quic-profile,
    number =    {draft-many-tiptop-quic-profile-00},
    type =      {Internet-Draft},
    institution =   {Internet Engineering Task Force},
    publisher = {Internet Engineering Task Force},
    note =      {Work in Progress},
    url =       {https://datatracker.ietf.org/doc/draft-many-tiptop-quic-profile/00/},
    author =    {Marc Blanchet},
    title =     {{QUIC Profile for Deep Space}},
    pagetotal = 11,
    year =      2025,
    month =     feb,
    day =       19,
}

@misc{rfc9000,
    series =    {Request for Comments},
    number =    9000,
    howpublished =  {RFC 9000},
    publisher = {RFC Editor},
    doi =       {10.17487/RFC9000},
    url =       {https://www.rfc-editor.org/info/rfc9000},
    author =    {Jana Iyengar and Martin Thomson},
    title =     {{QUIC: A UDP-Based Multiplexed and Secure Transport}},
    pagetotal = 151,
    year =      2021,
    month =     may,
}

@misc{PEP,
    series =    {Request for Comments},
    number =    3135,
    howpublished =  {RFC 3135},
    publisher = {RFC Editor},
    doi =       {10.17487/RFC3135},
    url =       {https://www.rfc-editor.org/info/rfc3135},
    author =    {Jim Griner and John Border and Markku Kojo and Zach D. Shelby and Gabriel Montenegro},
    title =     {{Performance Enhancing Proxies Intended to Mitigate Link-Related Degradations}},
    pagetotal = 45,
    year =      2001,
    month =     jun,
}

@INPROCEEDINGS{PEPsal,
  author={Caini, C. and Firrincieli, R. and Lacamera, D.},
  booktitle={Proc. 2006 IEEE 63rd Veh. Technol. Conf.}, 
  title={PEPsal: a Performance Enhancing Proxy designed for TCP satellite connections}, 
  year={2006},
  volume={6},
  number={},
  pages={2607-2611},
  doi={10.1109/VETECS.2006.1683339}}

@article{QUICL,
title = {QUICL: Disruption-tolerant networking via a QUIC convergence layer},
journal = {Comput. Commun},
volume = {241},
pages = {108247},
year = {2025},
issn = {0140-3664},
doi = {https://doi.org/10.1016/j.comcom.2025.108247},
author = {Markus Sommer and Artur Sterz and Markus Vogelbacher and Hicham Bellafkir and Bernd Freisleben},
}

@ARTICLE{Semantic-Aware,
  author={Gao, Ronghao and Li, Yue and Zhang, Qinyu and Yang, Zhihua},
  journal={IEEE Trans. Mob. Comput}, 
  title={Semantic-Aware Bundle Delivery in Space Disruption-Tolerant Networks via Cross-Layer Design on BP and LTP}, 
  year={2024},
  volume={23},
  number={9},
  pages={8742-8756},
  doi={10.1109/TMC.2024.3351428}}

@ARTICLE{HRBPRDS,
  author={Yang, Lei and Wang, Ruhai and Zhou, Yu and and others},
  journal={IEEE Trans. Veh. Technol.}, 
  title={Hybrid Retransmissions of Bundle Protocol for Reliable Deep-Space Vehicle Communications in Presence of Link Disruption}, 
  year={2021},
  volume={70},
  number={5},
  pages={4968-4983},
  doi={10.1109/TVT.2021.3072288}}

@ARTICLE{PLDTN,
  author={Alessi, Nicola and Caini, Carlo and de Cola, Tomaso and others},
  journal={IEEE Trans. Aerosp. Electron. Syst.}, 
  title={Packet Layer Erasure Coding in Interplanetary Links: The LTP Erasure Coding Link Service Adapter}, 
  year={2020},
  volume={56},
  number={1},
  pages={403-414},
  doi={10.1109/TAES.2019.2916271}}

@ARTICLE{ASCLBP,
  author={Zhou, Yu and Wang, Ruhai and Zhao, Kanglian and Burleigh, Scott C.},
  journal={IEEE Trans. Veh. Technol.}, 
  title={A Study of Cross-Layer BP/LTP Data Block Size in Space Vehicle Communications Over Lossy and Highly Asymmetric Channels}, 
  year={2020},
  volume={69},
  number={12},
  pages={16126-16141},
  doi={10.1109/TVT.2020.3042927}}

@article{Hybla,
author = {Caini, Carlo and Firrincieli, Rosario},
title = {TCP Hybla: a TCP enhancement for heterogeneous networks},
year = {2004},
issue_date = {September 2004},
publisher = {John Wiley \& Sons, Inc.},
address = {USA},
volume = {22},
number = {5},
issn = {1542-0973},
journal = {Int. J. Satell. Commun. Netw.},
month = sep,
pages = {547–566},
numpages = {20}
}

@article{BBR,
author = {Cardwell, Neal and Cheng, Yuchung and Gunn, C. Stephen and others},
title = {{BBR}: congestion-based congestion control},
year = {2017},
issue_date = {February 2017},
publisher = {Association for Computing Machinery},
address = {New York, NY, USA},
volume = {60},
number = {2},
issn = {0001-0782},
journal = {Commun. ACM},
month = jan,
pages = {58–66},
numpages = {9}
}

@ARTICLE{OAC-TCPCC,
  author={Masood, Arooj and Ha, Taeyun and Lakew, Demeke Shumeye and others},
  journal={IEEE Trans. Netw. Sci. Eng.}, 
  title={Intelligent TCP Congestion Control Scheme in Internet of Deep Space Things Communication}, 
  year={2023},
  volume={10},
  number={3},
  pages={1472-1486},
  doi={10.1109/TNSE.2022.3212534}}

@misc{MASQUE1,
    series =    {Request for Comments},
    number =    9484,
    howpublished =  {RFC 9484},
    publisher = {RFC Editor},
    doi =       {10.17487/RFC9484},
    url =       {https://www.rfc-editor.org/info/rfc9484},
    author =    {Tommy Pauly and David Schinazi and Alex Chernyakhovsky and others},
    title =     {{Proxying IP in HTTP}},
    pagetotal = 37,
    year =      2023,
    month =     oct,
}

@misc{MASQUE2,
    series =    {Request for Comments},
    number =    9298,
    howpublished =  {RFC 9298},
    publisher = {RFC Editor},
    doi =       {10.17487/RFC9298},
    url =       {https://www.rfc-editor.org/info/rfc9298},
    author =    {David Schinazi},
    title =     {{Proxying UDP in HTTP}},
    pagetotal = 16,
    year =      2022,
    month =     aug,
}

@INPROCEEDINGS{SMAQ,
  author={Kosek, Mike and Spies, Benedikt and Ott, Jörg},
  booktitle={Proc. IFIP Networking Conf.}, 
  title={Secure Middlebox-Assisted QUIC}, 
  year={2023},
  volume={},
  number={},
  pages={1-9},
  doi={10.23919/IFIPNetworking57963.2023.10186363}}

@ARTICLE{QUIRL,
  author={Michel, François and Bonaventure, Olivier},
  journal={IEEE/ACM Trans. Netw.}, 
  title={QUIRL: Flexible QUIC Loss Recovery for Low Latency Applications}, 
  year={2024},
  volume={32},
  number={6},
  pages={5204-5215},
  doi={10.1109/TNET.2024.3453759}}

@misc{KCP,
  author       = {{xtaci}},
  title        = {{kcptun: A Secure Tunnel based on KCP with N:M multiplexing and FEC}},
  howpublished = {\url{https://github.com/xtaci/kcptun}},
  month        = {Mar},
  year         = {2023},
  note = {Accessed: 26 August 2025}
}

@ARTICLE{StreamingCoding,
  author={Li, Ye and Zhang, Feifan and Wang, Jue and Quek, Tony Q. S. and Wang, Jiangzhou},
  journal={IEEE Commun. Lett.}, 
  title={On Streaming Coding for Low-Latency Packet Transmissions Over Highly Lossy Links}, 
  year={2020},
  volume={24},
  number={9},
  pages={1885-1889},
  doi={10.1109/LCOMM.2020.2989367}}

@article{PEPesc,
  title={PEPesc: a TCP performance enhancing proxy for non-terrestrial networks},
  author={Li, Ye and Chen, Liang and Su, Li and Zhao, Kanglian and Wang, Jue and Yang, Yongjie and Ge, Ning},
  journal={IEEE. Trans. Mob. Comput.},
  volume={23},
  number={4},
  pages={3060--3076},
  year={2023},
  publisher={IEEE}
}

@article{DecodingCost,
  title={On the Decoding Cost of Streaming Forward Erasure Correction Codes},
  author={Yu, Jianhao and Liu, Tengfei and Li, Ye and Ding, Feng and Wu, Sheng},
  journal={IEEE Commun. Lett.},
  volume={28},
  number={8},
  pages={1760--1764},
  year={2024},
  publisher={IEEE}
}

@misc{RFC5705,
    series =    {Request for Comments},
    number =    5705,
    howpublished =  {RFC 5705},
    publisher = {RFC Editor},
    doi =       {10.17487/RFC5705},
    url =       {https://www.rfc-editor.org/info/rfc5705},
    author =    {Eric Rescorla},
    title =     {{Keying Material Exporters for Transport Layer Security (TLS)}},
    pagetotal = 7,
    year =      2010,
    month =     mar,
}

@book{XQueue,
  author       = {Kleinrock, Leonard},
  title        = {Queueing Systems: Volume 1: Theory},
  publisher    = {Wiley‐Interscience},
  address      = {Hoboken, NJ, USA},
  year         = {1975},
  edition      = {1st},
  isbn         = {978-0471491101},
}

@ARTICLE{OFGE,
  author={Bioglio, Valerio and Grangetto, Marco and Gaeta, Rossano and Sereno, Matteo},
  journal={IEEE Commun. Lett.}, 
  title={On the fly gaussian elimination for LT codes}, 
  year={2009},
  volume={13},
  number={12},
  pages={953-955},
  doi={10.1109/LCOMM.2009.12.091824}}

@inproceedings{ACTIS,
  title={Advanced communications technologies in support of NASA mission},
  author={Miranda, F{\'e}lix A},
  booktitle={Proc. Eur. Conf. Antennas Propag.},
  number={GRC-E-DAA-TN53779},
  year={2018}
}

@INPROCEEDINGS{GEOPico,
  author={Jahandar, Saeid and Deutschmann, Jörg and Hielscher, Kai-Steffen and German, Reinhard},
  booktitle={Proc. 40th Int. Commun. Satell. Syst. Conf.}, 
  title={Performance of the QUIC transport protocol over geostationary satellite links}, 
  year={2023},
  volume={2023},
  number={},
  pages={222-227},
  doi={10.1049/icp.2024.0851}}

@ARTICLE{COV,
  author={Tsaoussidis, V. and Chi Zhang},
  journal={IEEE J. Sel. Areas Commun.}, 
  title={The dynamics of responsiveness and smoothness in heterogeneous networks}, 
  year={2005},
  volume={23},
  number={6},
  pages={1178-1189},
  doi={10.1109/JSAC.2005.845627}}

@ARTICLE{SFI,
  author={Tsaoussidis V and Chi Zhang},
  journal={IEEE J. Sel. Areas Commun.}, 
  title={The dynamics of responsiveness and smoothness in heterogeneous networks}, 
  year={2005},
  volume={23},
  number={6},
  pages={1178-1189},
  doi={10.1109/JSAC.2005.845627}}

\end{document}